\documentclass[12pt,a4paper,aps,preprint,preprintnumbers,superscriptaddress,nofootinbib]{revtex4-1}
\usepackage[english]{babel}
\usepackage[utf8]{inputenc}
\usepackage{graphicx}   
\usepackage{slashed}
\usepackage{epstopdf}
\usepackage{verbatim}   
\usepackage{color}      
\usepackage{subfigure}  
\usepackage{subfigure}
\usepackage{multirow}

\usepackage[colorlinks=true, pdfstartview=FitV, linkcolor=blue, citecolor=blue, urlcolor=blue]{hyperref}
\usepackage{float}
\restylefloat{table}
\usepackage{bm}
\usepackage[normalem]{ulem}
\usepackage{amsmath,amssymb}
\DeclareUnicodeCharacter{2212}{\textendash}

\def\n{\nonumber}

\definecolor{darkgreen}{rgb}{0,0.6,0}

\definecolor{orange}{rgb}{1,0.5,0}
\definecolor{blue}{rgb}{0,0,1}

\begin{document}

\title{Measuring Top Yukawa Coupling through $2\rightarrow 3$ VBS at Muon Collider }

\author{Junmou Chen}
\email{chenjm@jnu.edu.cn}

\affiliation{Jinan University, Guangzhou, Guangdong, China}

\author{Junhong Chen}
\email{2aron@stu2021@jnu.edu.cn}
\affiliation{Jinan University, Guangzhou, Guangdong, China}

\author{Wei He}
\email{weii20021121@stu2021.jnu.edu.cn}
\affiliation{Jinan University, Guangzhou, Guangdong, China}

\begin{abstract}
We study the measurement of top Yukawa coupling through $2\rightarrow 3$ VBS at future muon colliders, focusing on the lepton and semi-lepton channels of $\nu\nu t\bar th/Z$. First, analyzing  the partonic  amplitudes of $W_LW_L\rightarrow t\bar t h/Z_L$ and  simulating the full processes of $\nu\nu t\bar th/Z$ without decaying, we find they are highly sensitive to the anomalous top Yukawa $\delta y_t$. This sensitivity is enhanced by  selecting helicities  of the final $t\bar t$ and $Z$  to be $t_L\bar t_L+t_R\bar t_R$ and $Z_L$, which serves as the default and core setting of our analysis.
We then obtain the limits on $\delta y_t$  with this setting, giving $[-1.0\%, 1.1\%]$ for $\nu\nu t\bar th$ only and $[-0.36\%, 0.92\%]$ for $\nu\nu t\bar th$ and $\nu\nu t\bar tZ$ combined at $30$ TeV and  $1\sigma$.   Second,  we  proceed to analyze the processes after decaying and with background processes. To enhance the sensitivity  to $\delta y_t$, our settings  include  selecting the helicities of the final particles, as well as applying suitable cuts. However, we don't do  bin-by-bin analysis.    We obtain the limits on $\delta y_t$  for those channels at $10/30$ TeV and $1\sigma/2 \sigma$. The best limit is from the semi-lepton channel of $\nu\nu t\bar th$. With spin tagging efficiency at $\epsilon_s=0.9$,   it gives  $[-1.6\% , 1.8\%]$ at $1\sigma$ and $ [-2.4\%, 2.7\% ]$ at $2\sigma$ at $30$ TeV; $[-7.0\%, 6.7\%]$ at $1\sigma$ and $[-9.8\%, 9.8\%]$ at $2\sigma$ at $10$ TeV. 
\end{abstract}

\maketitle


\section{Introduction}

The top Yukawa coupling is one of the most important parameters in the standard model(SM), probably second only to Higgs self-couplings. In many new physics models, it's closely related to solution of the hierarchy problem, see for example\cite{KAPLAN1984187, DUGAN1985299, Bally:2022naz}.  Moreover, the CP phase  of top Yukawa could be related to the matter-anti-matter asymmetry of our universe\cite{Trodden:1998ym, Zhang:1994fb}.  Thus a precise measurement of top Yukawa is of great importance in the research of  particle physics today. This importance is further enhanced in light of the null result of  discovering supersymmetric particles or new particles in other new physics models\cite{Baer:2023uwo, ATLAS:2024zkx, ATLAS:2024vyj} at LHC.

Currently in LHC the top Yukawa coupling and its CP phase have been measured through $t\bar th$ and $th$ channels\cite{ATLAS:2023cbt,CMS:2020cga, CMS:2020djy}, albeit with large uncertainties. For example, the uncertainty of top Yukawa is still around $50\%$ at $2\sigma$(\cite{CMS:2020djy}).  This is to be expected, since hadron colliders are not designed for precision measurement due to its messy background.  Therefore, it's important to consider other options such as new colliders, among which  muon collider is a  particularly attractive candidate because of its advantages of both clean background and high energy.  Because of this advantage, future high energy muon collider has received considerable interests in the literature in recent years\cite{Delahaye:2019omf,Long:2020wfp,Buttazzo:2018qqp,Costantini:2020stv,Han:2020pif,Chiesa:2020awd}. As for  top Yukawa measurement in muon collider,  most research  in the literature so far focuses on the  $\nu\nu t\bar t$ channel, including for the top Yukawa coupling in \cite{Whisnant:1994fh, Henning:2018kys,  Maltoni:2019aot, Chen:2022yiu, Liu:2023yrb} and  its CP phase in \cite{Barman:2022pip}, which has the largest cross section for top production.  
 In \cite{Liu:2023yrb}, the authors found $\delta y_t$ can be constrained up to $1.5\%(1\sigma)$ at  10 TeV muon collider with  $10\  \text{ab}^{-1}$ luminosity, while in \cite{Chen:2022yiu} the precision of $y_t$ at $2\sigma$ is projected to be $4.5\%(1.4\%)$ at $10\ \text{TeV}(30\ \text{TeV})$.   Besides $\nu\nu t\bar t$, there are also a few studies on other channels, see \cite{Cassidy:2023lwd} and \cite{Chen:2022yiu}. In the former, $\nu\nu t\bar th$ and $lvtbh$ are studied along with $\nu\nu t\bar t$ in measuring the CP phase.  In the latter, the measurement of top Yukawa  is studied using $\nu\nu t\bar th$, along with  $\nu\nu t\bar t$ as mentioned above. The limits on $\delta y_t$ at $2\sigma$ level  from $\nu\nu t\bar th$ were obtained to be around $5.5\%(0.69\%)$ at $10(30)$ TeV. In both cases cited above(\cite{Liu:2023yrb, Chen:2022yiu}),  the analysis is done with the final states($\nu \nu t\bar t$/$\nu \nu t\bar th$) directly without decaying and hadronization, with no or only superficial background analysis.  Moreover, they both use bin-by-bin method in data analysis. Thus their results are highly idealized.

The focus on the $\nu\nu t\bar t$ channel is understandable,  considering its large cross section. However,  it's also important to consider and further explore other options. In this paper, we  study the potential of using $2\rightarrow 3$ vector boson scattering(VBS) processes  ($VV\rightarrow t\bar t h/V_L$) to measure the top Yukawa coupling, including but not limited to the $\nu\nu t\bar th$ channel. 
The motivation can be simply explained in the following. By taking Goldstone equivalence,  the $2\rightarrow 3$ VBS amplitude with longitudinal vector bosons has a contact diagram from a dim-6 SMEFT operator  ($\mathcal O_6$ in the case of $\varphi \varphi \rightarrow \varphi\varphi h$($V_LV_L\rightarrow V_LV_L h$),  $\mathcal O_{t\Phi}$(\cite{Dedes:2017zog}) in the case of $\varphi \varphi\rightarrow t\bar th/\varphi(V_LV_L\rightarrow t\bar t h/V_L)$). In high energy, the contact diagram dominates the amplitude. Since  $\mathcal O_{t\Phi}$ ---- the  SMEFT operator that gives the contact diagram ----  also gives rise to an anomalous top Yukawa coupling $\delta y_t$,  we conclude that the VBS process would be highly sensitive to $\delta y_t$. This situation is  in analogue to  the measurement of Higgs self-couplings through $\nu\nu V_L V_Lh$ channels as done in \cite{Chen:2021rid, Chen:2021pqi}. Crucially, the above analysis would lead to the conclusion that the helicities of final $t\bar t$ have to be $t_R\bar t_R$ or $t_L\bar t_L$ in high energy. The analysis will be explained in more details in the next section. 


In this paper we won't touch on the top CP phase, but focus on the anomalous top Yukawa coupling $\delta y_t$ only.  There are two goals of our paper: First we intend to find a simple method to achieve high significance, which do not rely on bin-by-bin analysis used in \cite{Chen:2021rid, Chen:2021pqi} and etc. Since it is very time consuming and  more appropriate for optimization after a simple method has been found to achieve high sensitivity. Second, we wish to  obtain more reliable estimations for $\delta y_t$. Our central setting to achieve the first goal is to select helicities of final states ($t\bar t $ and $V$ before decaying) to be $t_R\bar t_R +t_L\bar t_L$ and $V_L$. For the second goal, we will carry out a more complete analysis by  decaying the final states $t\bar t$ and $h/V$ and simulating background processes along with the signal.  We then analyze the data with the combination of  selecting final states' helicities and some simple cuts. Finally we choose to do the analysis at  the center-of-mass energy(c.m.)  of $\sqrt{s}=10, 30 $ TeV.


The rest of the paper is organized as following:

In Section  (\ref{sec:part_proc}), we will study the processes at the level of the VBS processes and hard processes before decaying. Focusing on $\nu\nu t\bar th/Z$ after listing all $2\rightarrow 3$ VBS related to top production, we first analyse  the amplitude of $W_LW_L\rightarrow t\bar th/Z_L$ with $\mathcal O_{t\Phi}$.  After that, we obtain the statistic significance of $\nu\nu t\bar th/Z$ with and without selecting helicities as comparison, within a range of c.m. energies.  We finally obtain the limit of $\delta y_t$ for $\nu\nu t\bar th/Z$ without decaying and background analysis, as an approximation to the limits of all the data when made use of sufficiently.


In Section  (\ref{sec:vvtth}), we study $\nu\nu t\bar th$ at the lepton channel and the semi-lepton channel respectively. We generate all important background processes, then obtain the cuts to reduce the background relative to signal. We then obtain the limits on $\delta y_t(c_{t\phi})$ at $30$ TeV and $10$ TeV, unless the statistical significance is too small. 

In Section (\ref{sec:vvttz}), we study $\nu\nu t\bar t Z$ at the lepton channel and the semi-lepton channel, following the same approach in Sec.(\ref{sec:vvtth}). 

In Conclusion (\ref{sec:con}), we summary and discuss our results, compare them with the literature, and finally discuss future research directions.



\section{First Analysis of Parton and Hard Processes }
\label{sec:part_proc}
\subsection{Partonic Processes: $VV\rightarrow t\bar{t}V/h$  }

In  high energy  muon colliders, vector boson scatterings become the dominant channels in producing top quarks.  For $2\rightarrow 3$ VBS we have the following list of processes that are sensitive to   the anomalous top Yukawa coupling:
\begin{eqnarray}\label{eq:processes}
   && W^+ W^-/ZZ\rightarrow t\bar{t} h,\ \ \  W^+ W^-/ZZ \rightarrow t \bar{t} Z,  \ \ \ W^+W^-/ZZ\rightarrow t \bar{b} W^-(/ \bar{t}bW^+) \\
    &&   W^{\pm} Z \rightarrow t \bar{t} W^{\pm}, \ \ \ W^{\pm}Z\rightarrow t\bar{b}(/ b\bar{t}) Z, 
\end{eqnarray}
$WW$ scatterings correspond to $\mu\mu\rightarrow \nu\bar{\nu} X$ in a muon collider, $ZZ$ scatterings correspond to $\mu\mu \rightarrow \mu\mu X$, while $WZ$  scatterings correspond to $\mu\mu\rightarrow \mu\nu X$.  In this paper we focus on $WW$ scatterings only, since in our experience their  cross sections dominate over the others.  Moreover, $W^+W^-/ZZ\rightarrow t \bar{b} W^-(/ \bar{t}bW^+)$ has the same final states as $WW\rightarrow t\bar{t}, t/\bar t\rightarrow bW$, thus should be analyzed together with $2\rightarrow 2$ VBS.  So it won't be part of our study here either. Therefore the processes we study in this paper are 
\begin{equation}
W^+ W^-\rightarrow t\bar{t} h,\ \ \  W^+ W^-\rightarrow t \bar{t} Z,
\end{equation}

The general Lagrangian terms that give the top Yukawa coupling can be written as following:
 \begin{equation}\label{eq:anomalous_top_yukawa_1}
    \mathcal{L} \supset   -\frac{y_t^{\text{SM}}}{\sqrt{2}}\kappa_t\bar{t}(\cos\delta+i\gamma_5\sin\delta )th
\end{equation}  
$y_t^{\text{SM}}=\sqrt{2}m_t/v$ is the SM value of top Yukawa, $\kappa_t$ and CP phase $\delta$ are free parameters that measure the deviation from the SM. In this paper we're not interested in CP violation, so $\delta$ is set to be $0$ from now on. We further split $\kappa_t$ as $\kappa
_t=1+\delta y_t$, the Lagrangian then becomes
\begin{equation}\label{eq:anomalous_top_yukawa_2}
\mathcal{L} \supset   -\frac{y_t^{\text{SM}}}{\sqrt{2}}(1+\delta y_t)\bar{t}th
\end{equation}

However, Eq.(\ref{eq:anomalous_top_yukawa_1}) and Eq.(\ref{eq:anomalous_top_yukawa_2}) do not include all relevant physics, since the Higgs boson in the SM belongs to a $SU(2)$ doublet, along with other would-be Goldstone bosons $\varphi^{\pm}, \varphi^0$.  To include them, we need to consider the gauge invariant SM Lagrangian:
\begin{equation}\label{eq:lag_top_SM}
 \mathcal{L} \supset  -y_t^{\text{SM}} \overline Q t_R\tilde\Phi+ \text{h.c.}
\end{equation}
in which $Q=(t_L, b_L)^T$, $\Phi=(\varphi^+, \frac{1}{\sqrt{2}}(h+i\varphi^0))^T$ and $\tilde\Phi_a=\epsilon_{ab}\Phi_b$. To further accommodate the anomalous  top Yukawa coupling gauge invariantly, we can add the SMEFT operator 
\begin{equation}\label{eq:lag_top_O6}
\mathcal{O}_{t\phi}=\frac{c_{t\phi}}{\Lambda^2}\overline Qt_R\tilde{\Phi}(\Phi^\dagger\Phi)+\text{h.c.}
\end{equation}
The energy scale $\Lambda$ is taken to be $1$ TeV in this paper, we can also define $C_{t\phi}\equiv \frac{c_{t\phi}}{\Lambda^2}$ to absorb $\Lambda$ into $c_{t\phi}$.

In unitarity gauge, Eq.(\ref{eq:lag_top_SM}) and Eq.(\ref{eq:lag_top_O6}) combined gives the $tth$ coupling in Eq.(\ref{eq:anomalous_top_yukawa_1}) with  
\begin{equation}\label{eq:delyt_ctp}
\Delta y_t \equiv \delta y_t\cdot y_t^{\text{SM}}=-\frac{c_{t\phi}v^2}{\Lambda^2}.
\end{equation}

Although in principle  all physical information can be extracted in unitary gauge, it's usually not the most convenient and physically clarifying approach, because of the large cancellation between Feynman diagrams.   Instead, we choose a gauge  with Goldstone bosons (e.g. Feynman gauge) and take Goldstone equivalence, which simplify the analysis significantly.

Let's analyze the amplitude of $W_L^+W_L^-\rightarrow t \bar t h$ as an  example, which  can be approximated by $\varphi^+ \varphi^- \rightarrow t \bar t h$ in high energy.  Notice $\mathcal O_{t\Phi}$(Eq.(\ref{eq:lag_top_O6})) gives a 5-point contact vertex/diagram: $\lambda_{\varphi\varphi tth}=i\frac{c_{t\phi}v^2}{\sqrt{2}\Lambda^2}$. In the high energy limit, other diagrams with one or two propagators are suppressed by $\mathcal O(\frac{1}{E^2})$, while only the contact diagram remains constant\footnote{Soft/collinear  logarithm can elevate this suppression to certain extent, but don't change the qualitative picture.}.
The BSM amplitude with $\lambda_{\varphi\varphi tth}$  is therefore  approximately:  
\begin{equation}
    \mathcal A_{\text{BSM}} \simeq i\frac{c_{t\phi}v^2}{\sqrt{2}\Lambda^2} + \mathcal O(\frac{m_W^2}{E^2}).
\end{equation}
In contrast, the SM amplitude is suppressed by energy:
\begin{equation}
    \mathcal A_{\text{SM}} \simeq \mathcal O(\frac{m_W^2}{E^2}).
\end{equation}
Comparing BSM and SM amplitudes indicates high sensitivity of $\nu\nu t\bar th$ to $\delta y_t$, giving justification for our choice of processes in this paper.  
The similar conclusion can be drawn for $WW\rightarrow t \bar t Z$. Of course, in order for Goldstone equivalence to apply, $Z$ has to be longitudinal so that $Z_L \sim \phi^0$ in high energy.

There remains a subtlety with the analysis above, involving the helicities of $t\bar t$. Notice $\mathcal O_{t\Phi}$ gives the chiral structure of  $\bar t_L t_R$ or $\bar t_R t_L$, similar to the SM  $tth$ coupling. In high energy limit for the final states, $t_{L/R}$ in chirality corresponds to $t_{L/R}$ in helicity only, while $\bar t_{L/R}$ in chirality corresponds to $\bar t_{R/L}$ in helicity only. Therefore, only $\bar t_{R}t_{R}$ or $\bar t_L t_L$ (in helicity)  contributes to the contact diagram $\lambda_{\varphi^+\varphi^- t th}(\lambda_{\varphi^+\varphi^- t t \varphi^0})$ of $\varphi^+\varphi^-\rightarrow \bar t t h(\varphi^+ \varphi^- \rightarrow  \bar t t\varphi^0)$. 

 The finding above   indicates we can significantly enhance the sensitivity to the anomalous top Yukawa by selecting the helicities of the final $t\bar t$ and $Z$. We can make a rough estimation of the enhancement  to  statistical significance $\mathcal S$, as defined by  
 \begin{equation}\label{eq:signf_apprm}
 \mathcal S \simeq \frac{S}{\sqrt{B}}= \frac{\sqrt{\mathcal L}\cdot \sigma^{\text{sig}}}{\sqrt{\sigma^{\text{bkg}}}} 
 \end{equation}
 with $S$ and $B$  being event numbers of signal and background respectively,  $\mathcal L$ being integrated luminosity.  For $\nu\nu tth$, the ratio of the cross section of summing over all spins to that of  $t_L\bar t_L + t_R\bar t_R$ is roughly $2:1$; while for $\nu\nu ttZ$, the ratio of full cross section to that of $t_L\bar t_LZ_L+ t_R\bar t_RZ_L$ is around $5:1$.  Then the  ratio of statistic significance from selecting helicity to not selecting helicity is 
 \begin{eqnarray}\label{eq:signf_ratio}
\nu\nu tth:&&\ \   \frac{{\mathcal S}_{\text{spin}}}{{\mathcal S}_{\text{no-spin}}}= \sqrt{\frac{\sigma^{\text{bkg}}_{\text{no-spin}}}{\sigma^{\text{bkg}}_{\text{spin}}}} = \sqrt{2}\simeq 1.4\\
\nu\nu ttZ:&&\ \   \frac{{\mathcal S}_{\text{spin}}}{{\mathcal S}_{\text{no-spin}}}= \sqrt{5} \simeq 2.23,  \nonumber
 \end{eqnarray}
giving $40\%$ enhancement for $\nu\nu tth$ and $120\%$ enhancement for $\nu\nu ttZ$. Notice the dependence on $\sigma^{\text{sig}}$ and $\mathcal L$ cancels out.

\subsection{Analysis of $\nu\nu t\bar th/\nu\nu t\bar tZ$ without Decaying }
\label{sec:parton_signf}

To have more accurate and general estimations, we use Madgraph5\cite{Alwall:2014hca}(version 3.4.1) to generate the total cross sections of $\nu\nu t\bar th$ and $\nu\nu t\bar tZ$ with different helicity combinations, and without decaying or considering background processes. The BSM processes are generated with SMEFTatNLO(\cite{Degrande:2020evl}) package at LO mode, $c_{t\phi}$ is set at $c_{t\phi}=2$ and $\Lambda = 1$ TeV. The c.m energy is scanned from $10$ TeV to $30$ TeV.  For the statistical significance $\mathcal S$ we use the following formula:
\begin{eqnarray}
    \mathcal S = \sqrt{2\left[(S+B)\cdot \text{ln}\left(1+\frac{S}{B}\right)-S\right]}
    \label{eq:signficance}
\end{eqnarray}
The signal ``S" is \begin{equation}
\label{S}
    S=(\sigma^{\text{sig}}_{\text{bsm}}-\sigma^{\text{sig}}_{\text{sm}})\cdot \mathcal{L}
\end{equation}the background ``B" is \begin{equation}
\label{B}
    B=\sigma^{\text{bkg}}\cdot \mathcal{L}.
\end{equation} 
Here ``sig" signifies the processes $\mu\mu\rightarrow \nu\bar\nu t_R\bar t_R(/t_L\bar t_L) h $ and/or $\mu\mu\rightarrow \nu\bar\nu t_R\bar t_R(/t_L\bar t_L)  Z_L $, $``\text{bkg}"$ signifies all other processes involved.

We obtain the integrated luminosity by using the following formula
\begin{equation}
\label{eq:lumin}
\mathcal{L}= \left(\frac{\sqrt{s}}{10\ \text{TeV}}\right)^2\cdot  10\ \text{ab}^{-1}
\end{equation}
which for example gives $10, 90 \ \text{ab}^{-1}$ for  $\sqrt{s}=10, 30\ \text{TeV}$ respectively.

\begin{figure}[t]\textbf{}
    \centering
    \includegraphics[width=75mm]{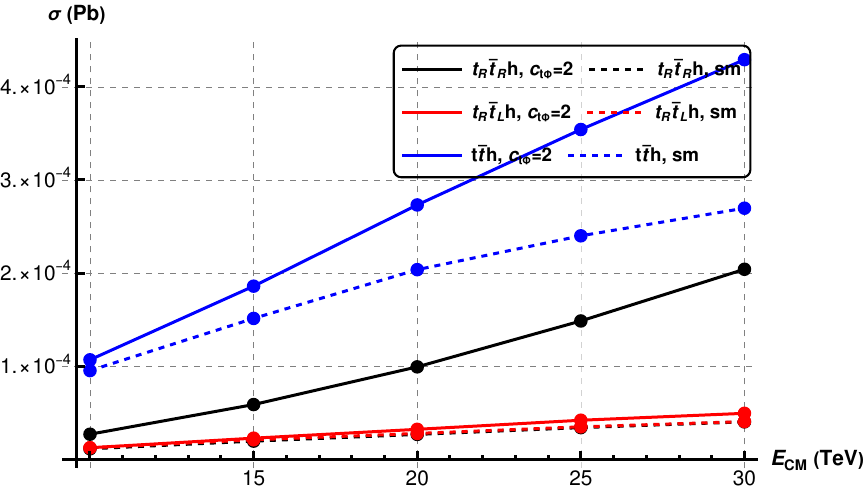}
    \hspace{0.01\textwidth}
    \includegraphics[width=75mm]{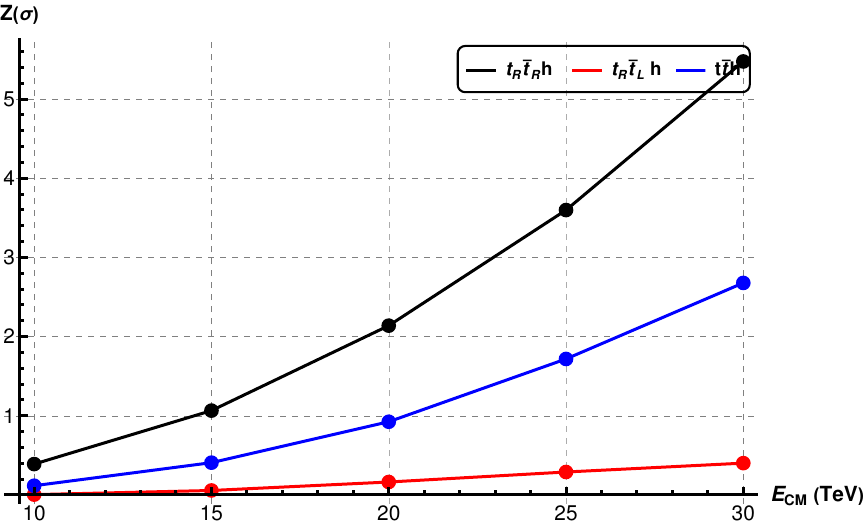}     
   \caption{The figure on the left shows cross sections of $\mu\mu\rightarrow \nu \nu t \bar t h$ at $c_{t\phi}=0$(SM) and $c_{t\phi}=2$(BSM), with helicity combinations of   $\nu\nu t_R\bar t_R h+ \nu\nu t_L\bar t_Lh$(denoted by $t_R\bar t_R h$), $\nu\nu t_R\bar t_L h+ \nu\nu t_L\bar t_R h$(denoted by $t_R\bar t_L h$) and $\nu\nu t\bar th$(denoted by $t\bar th$). No cuts are applied. The figure on the right shows the corresponding statistical significance of       $t_R \bar t_R h$, $t_R\bar t_L h$ and $t\bar t h$.  }
    \label{fig:sig_hel_vvtth}
\end{figure}

\begin{figure}[t]\textbf{}
    \centering
    \includegraphics[width=75mm]{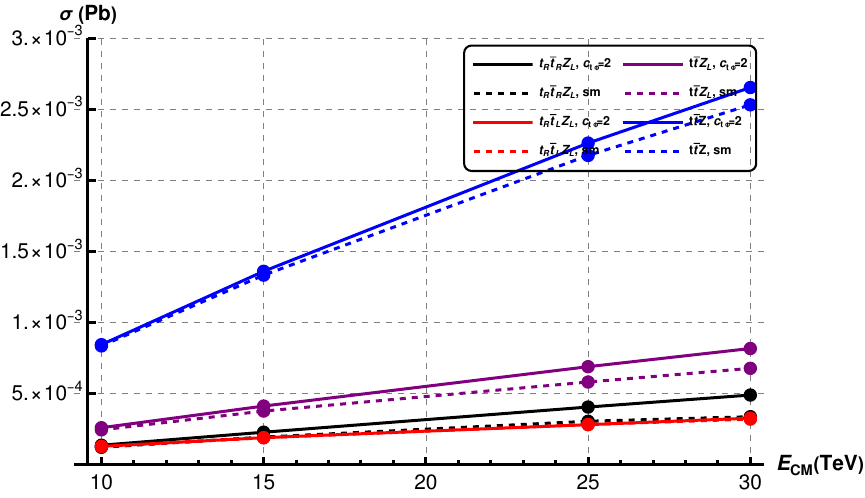}
    \hspace{0.01\textwidth}
    \includegraphics[width=75mm]{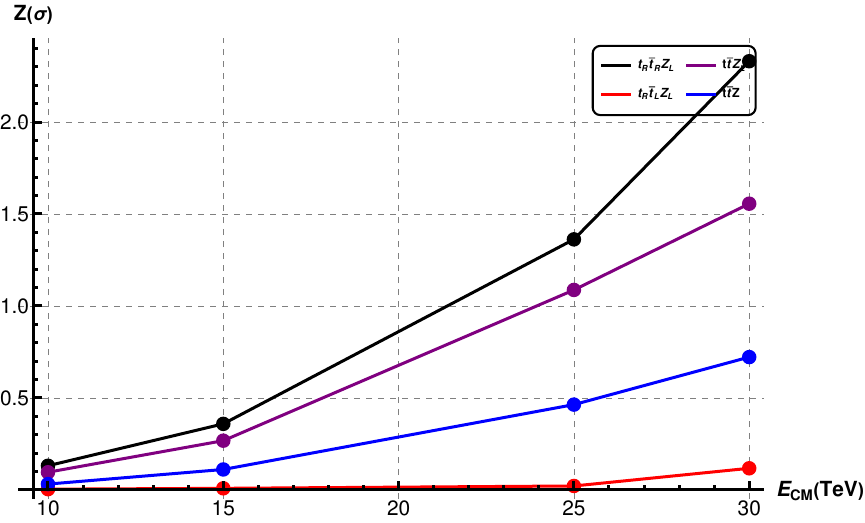} 
   \caption{The figure on the left shows cross sections of $\mu\mu\rightarrow \nu\nu t \bar t h$ at $c_{t\phi}=0$(SM) and $c_{t\phi}=2$(BSM), with helicity combinations of   $\nu \nu t_R\bar t_RZ_L+ \nu \nu t_L\bar t_LZ_L$(denoted by $t_R\bar t_R Z_L$), $\nu \nu t_R\bar t_L Z_L+\nu\nu t_L\bar t_R Z_L$(denoted by $t_R\bar t_L Z_L$), $\nu \nu t\bar tZ_L$(denoted by $t\bar tZ_L$) and  $\nu \nu t\bar tZ$(denoted by $t\bar tZ$). No cuts are applied. The figure on the right shows the statistical significance of $t_R\bar t_RZ_L$, $t_R\bar t_L Z_L$, $t\bar tZ_L$ and $t\bar tZ$. 
    }
    \label{fig:sig_hel_vvttz}
\end{figure}

The results are shown in Fig.(\ref{fig:sig_hel_vvtth}) for $\nu\nu t\bar th$ and Fig.(\ref{fig:sig_hel_vvttz}) for $\nu\nu t\bar t Z$. In Fig.(\ref{fig:sig_hel_vvtth}), we show the cross sections and significance of $\nu\nu t_R\bar t_Rh+\nu\nu t_L\bar t_Lh$, $\nu\nu t_R\bar t_Lh+\nu\nu t_L\bar t_Rh$ and $\nu\nu t\bar th$ without selecting helicities of $t\bar t$. In Fig.(\ref{fig:sig_hel_vvttz}), we show the cross sections and significance of $\nu\nu t_R\bar t_RZ_L+\nu\nu t_L\bar t_LZ_L$, $\nu\nu t_R\bar t_LZ_L+\nu\nu t_L\bar t_RZ_L$,  $\nu\nu t\bar tZ_L$ and $\nu\nu t\bar tZ$. In both cases, the results in the figures confirm the estimates in  Eq.(\ref{eq:signf_ratio}) qualitatively, i.e. selecting helicities for $t\bar t$ or/and $Z$ significantly enhances the statistical significance. Quantitatively however, there are some differences. The effects on $\nu\nu t\bar th$ are larger than naively expected, giving enhancement of close to $100\%$ for $\nu\nu t_R\bar t_R h+\nu\nu t_L\bar t_L h$ relative to $\nu\nu t\bar th$; similarly, the enhancement for $\nu\nu t_R\bar t_R Z_L+ \nu\nu t_L\bar t_L Z_L$ to $\nu\nu t\bar tZ$ is around $300\%$,   with the enhancement of $\nu\nu t\bar t z$ largely coming from selecting $Z_L$.  Based on the analysis above, we will specify the helicities of $t\bar t$ and $Z$ to be $t_R\bar t_R+ t_L\bar t_L$ and $Z_L$ respectively in simulation and analysis for the remainder of this paper. 

Finally, to have an understanding on how  $\delta y_t$ could be  constrained if all the data from $\nu\nu t\bar t h$ and $\nu\nu t\bar t Z$ are sufficiently made use of.  After some simple cuts and selecting helicities for the final states to be $t_R\bar t_R+t_L\bar t_L$ and $Z_L$,  we obtain the constraints on $\delta y_t$  for $\nu \nu t\bar t h$, $\nu \nu t\bar tZ$ and $\nu \nu t\bar th+vvt\bar tZ$ respectively without decaying at $1\sigma$ and $2\sigma$.  The results are listed in Table(\ref{tab:constrt_nodecay}). Not surprisingly,  $\nu\nu t\bar th$ gives more stringent limits than $\nu\nu t\bar t Z$, while the limits from $\nu\nu t\bar t h +\nu\nu t\bar tZ$ are more stringent than both of them.  As an example,  the $1\sigma$ limit from $\nu\nu t\bar t h$ at $30$ TeV is $[-1.0\%, 1.1\%]$, the same limit  from $\nu\nu t\bar t Z$ is  $[-0.43\%, 1.2\%]$, that from $\nu\nu t\bar th +\nu\nu t\bar tZ$ is $[-0.36\%, 0.92\%]$. The results are comparable to ref.(\cite{Chen:2022yiu}) in which the results are obtained with bin-by-bin analysis. The results from $\nu\nu t\bar t Z$, which have been neglected so far in the literature,  are only slightly worse than $\nu\nu t\bar t h$, making the combined results look particularly promising. 

\begin{table}[]
\centering
\begin{tabular}{|c|c|c|c|}
\hline  
\multirow{2}{*}{Channels}      & \multirow{2}{*}{$E_{cm}$}     &    \multicolumn{2}{|c|}{Limits of $\delta y_t$}                  \\
\cline{3-4} 
                               &                               &      $1\sigma$   &  $2\sigma$  \\
\hline 
\multirow{2}{*}{$\nu \nu t\bar t h$} &   30 TeV                      &$[-1.0\%, 1.1\%]$&  $[-1.1\%, 1.2\%]$  \\
\cline{2-4}
                               &    10 TeV                     & $[-3.5\%, 3.1\%]$ &  $[-4.9\%, 4.3\%]$\\
 \hline
\multirow{2}{*}{$\nu \nu  t\bar tZ$}  &    30 TeV                     &$[-0.43\%, 1.2\%]$&  $[-0.75\%, 1.5\%]$ \\
\cline{2-4} 
                              &    10 TeV                      &  $[-2.8\%, 5.2\%]$&  $[-4.4\%, 6.7\%]$\\
\hline 
\multirow{2}{*}{$\nu \nu  t\bar th+ 
\nu\nu t\bar t Z$}                  &    30 TeV                     & $[-0.36\%, 0.92\%]$ & $[-0.73\%, 1.2\%]$\\
\cline{2-4} 
                             &    10 TeV                       & $[-2.6\%, 3.1\%]$ & $[-3.7\%, 4.3\%]$ \\
\hline 
\end{tabular}
\caption{The constraints on $\delta y_t$ for $\nu\nu t\bar th$, $\nu\nu t\bar tZ$ and $\nu\nu t\bar th+\nu\nu t\bar tZ$ respectively without decaying. The helicities of $t\bar t$ are selected to be $t_R\bar t_R+t_L\bar t_L$, the helicity of $Z$ to be $Z_L$. To $\nu\nu t\bar t h$ we applied the cuts of $p_T(t) > 200$ GeV and $p_T(h) > 100$ GeV;  to $\nu\nu t\bar tZ$ we applied the cuts of  $p_T(t) > 200$ GeV,  $p_T(Z) > 250 $ GeV. } 
\label{tab:constrt_nodecay}
\end{table}

\pagebreak

\section{Full Analysis of $\nu\nu t\bar t h$}
\label{sec:vvtth}

In this section, we carry out a detailed  analysis on the lepton channel $l^+l^-2b+\text{MET}$ and the semi-lepton channel $l^\pm jj4b+\text{MET}$,  both of which  come mainly from   $\nu\nu t\bar t h$, with minor contribution from $\nu\nu t\bar tZ$.  Same as in parton processes, we simulate  with hard processes with Madgraph5$\_3\_4\_1$ and SMEFTatNLO\_LO. We also decay heavy particles with Madspin. However, since  muon colliders are not built yet, we deem it premature to do collider analysis, thus we don't implement parton shower and hadronization in our simulation. Nonetheless we believe simulation and analysis on the parton level is enough to obtain relatively reliable results.  To take into account of those effects, we set   $b$  tagging efficiency to be $0.9$.  We will also scan the tagging efficiencies of $Z_L$ and $t_R\bar t_R/t_L\bar t_L$ in the analysis to see how our results varies with taggin efficiencies. 
For the other settings we follow Sec.(\ref{sec:parton_signf}), the formula of statistical significance is given by Eq.(\ref{eq:signficance}). For c.m. energy and integrated luminosity, we use Eq.(\ref{eq:lumin}) choose $10\  \text{ab}^{-1}$ at $10$ TeV and $90 \ \text{ab}^{-1}$ at $30$ TeV as two benchmark points.

\subsection{Lepton Channel}

In this subsection  we  will study  the channel $l^+l^-4b+\text{MET}$  with $l=e,\mu$ and $b=b/\bar b$ of which the dominant  contribution comes from   $\nu\nu t\bar th$  process, but  minor contribution also come from $\nu\nu t\bar tZ$.  It's obtained from the hard process  through $t\rightarrow b W^+\rightarrow b l^+ \nu_l$, $\bar{t}\rightarrow \bar{b} W^-\rightarrow \bar b l^-\bar{\nu}_l$, $h/Z\rightarrow b \bar{b}$.   When analyzing data, we will always include contributions from $\nu\nu t\bar th$, but also  include $\nu\nu t\bar tZ$ when it's beneficial to do so. 

The main background processes are
\begin{eqnarray}
    &&\mu\mu\rightarrow \nu\bar \nu t_R\bar{t}_Rh/Z_L(\text{SM}) \rightarrow 2\nu2\bar\nu l^+l^- 4b\n\\
    &&\mu\mu\rightarrow t_R\bar t_R h \rightarrow \nu\bar\nu l^+l^- 4b\n\\
    &&\mu\mu\rightarrow \nu\bar \nu Z_LZ_LZ_L \rightarrow  \nu\bar \nu l^+l^-4b \n\\
    &&\mu\mu\rightarrow \nu\bar \nu Z_LZ_Lh \rightarrow  \nu\bar \nu l^+l^-4b\\
    &&\mu\mu\rightarrow \nu\bar \nu Z_Lhh \rightarrow  \nu\bar \nu l^+l^-4b\n\\
    &&\mu\mu\rightarrow l^+\nu_lW^-Z_LZ_L(/l^-\bar\nu_l W^+Z_LZ_L) \rightarrow \nu\bar\nu l^+l^-4b\n\\
    &&\mu\mu\rightarrow l^+\nu_lW^-Z_Lh(/l^-\bar\nu_l W^+Z_Lh) \rightarrow \nu\bar\nu l^+l^-4b\n\\
    &&\mu\mu\rightarrow l^+\nu_lW^-hh(/l^-\bar\nu_l W^+hh) \rightarrow \nu\bar\nu l^+l^-4b\n
\end{eqnarray}
with $t_R\bar t_R$ denoting $t_R\bar t_R+t_L\bar t_L$. 
There are also other background processes, such as $t\bar t Z$, $ZZZ$ and etc, but their cross sections are all negligible compared with those listed above.  The largest contribution to background is   $\nu\bar \nu t\bar{t}h/Z$ from SM.

\subsubsection{30 TeV}

We first study $30$ TeV for the lepton channel from $\nu\nu t\bar th$. For all the processes, we implement a set of preliminary cuts:
\begin{equation}
    p_T(b) > 20 \ \text{GeV}, \ \ \  p_T(l) > 30 \ \text{GeV}
\end{equation}
with $b = b /\bar{b}$ and  $l = l^+ /l^-$. 
We emphasize again that the crucial setting of our analysis is specifying helicities of final particles.  To accommodate collider environment, we will scan over different values of spin tagging efficiencies of $Z$ and $t/\bar t$ from $1$ to $0.6$, in doing analysis. 

\begin{figure}
    \centering
    \includegraphics[width=80mm]{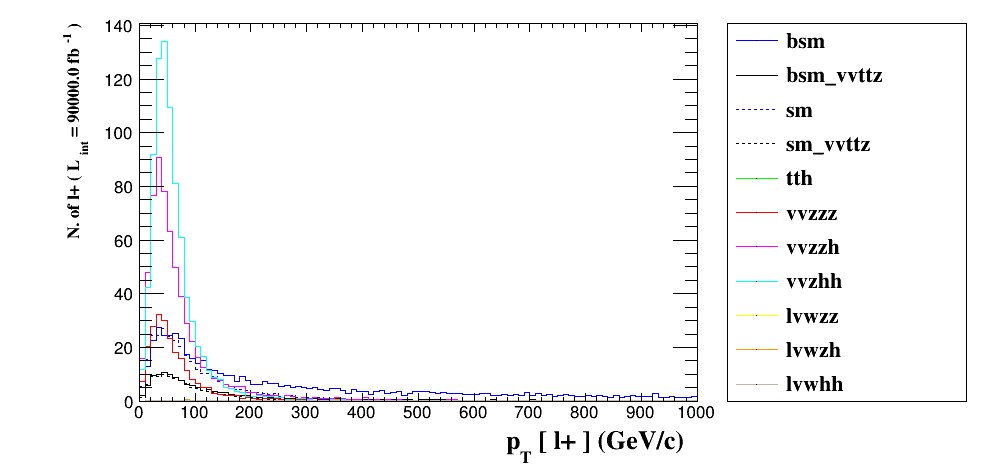}
    \includegraphics[width=80mm]{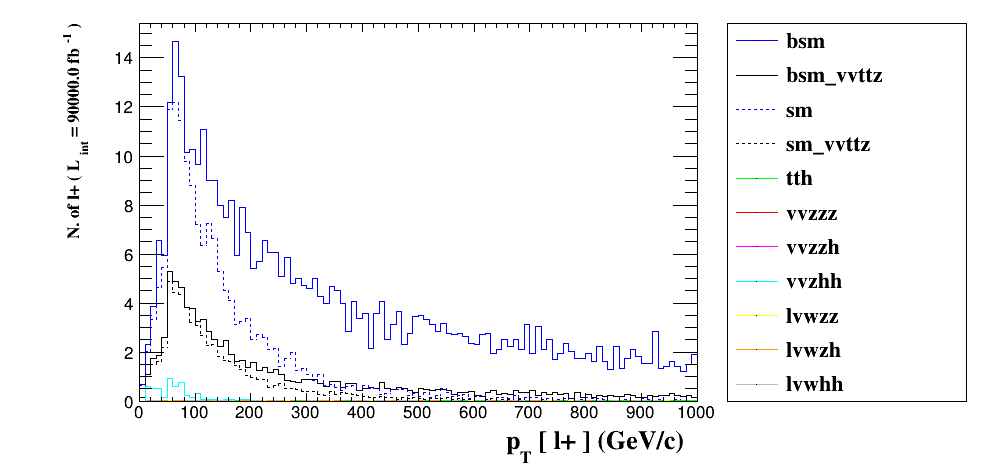}
    \includegraphics[width=80mm]{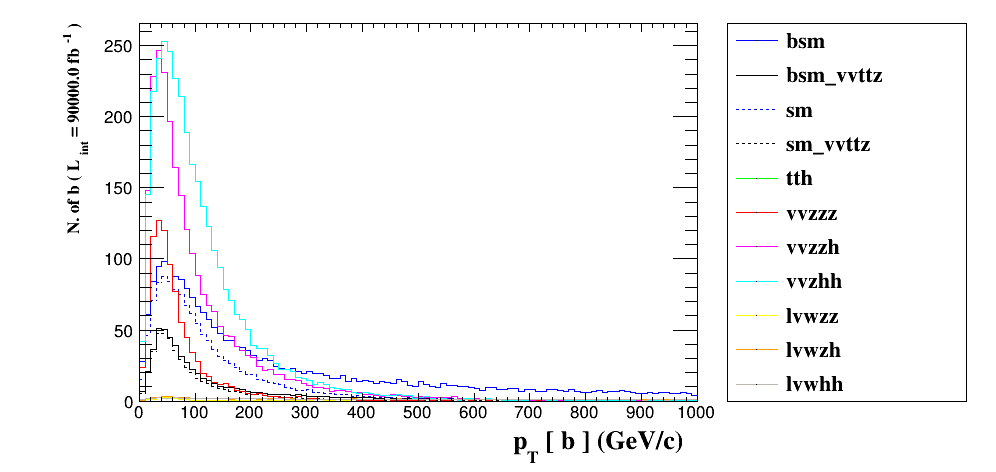}
    \includegraphics[width=80mm]{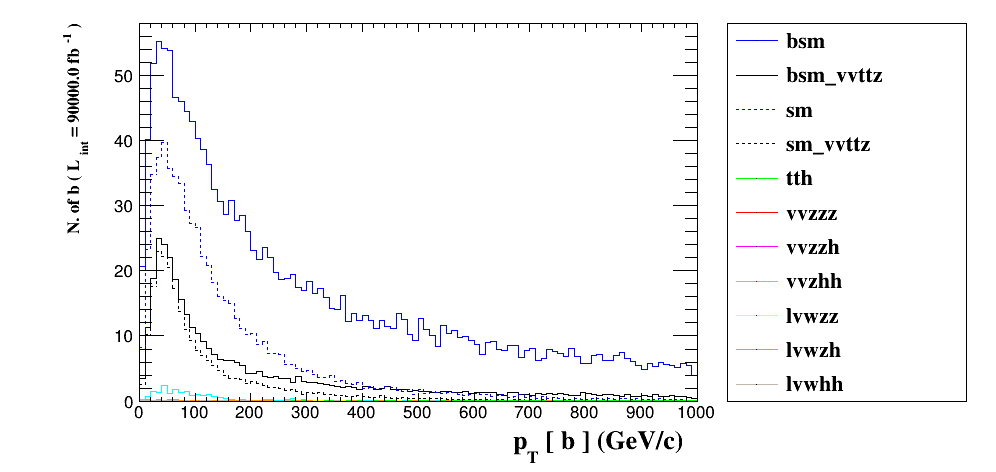}
    \includegraphics[width=80mm]{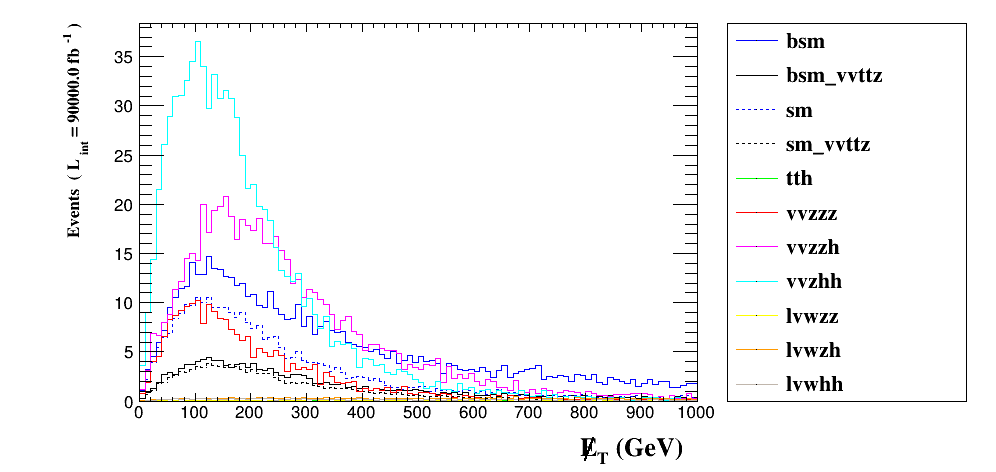}
    \includegraphics[width=80mm]{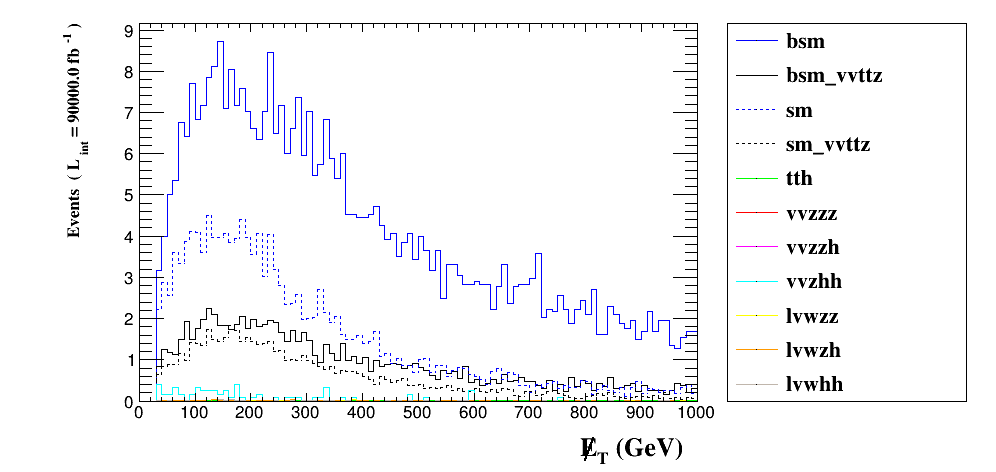}
    \includegraphics[width=80mm]{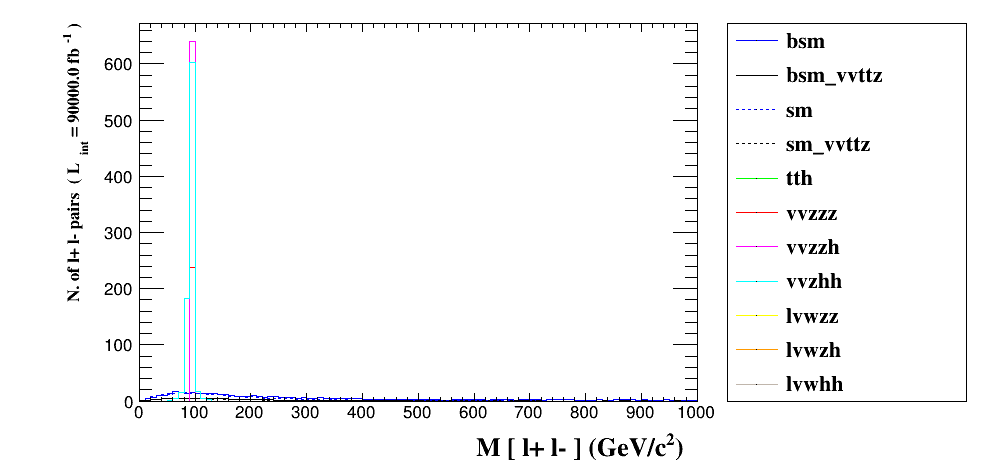}
    \includegraphics[width=80mm]{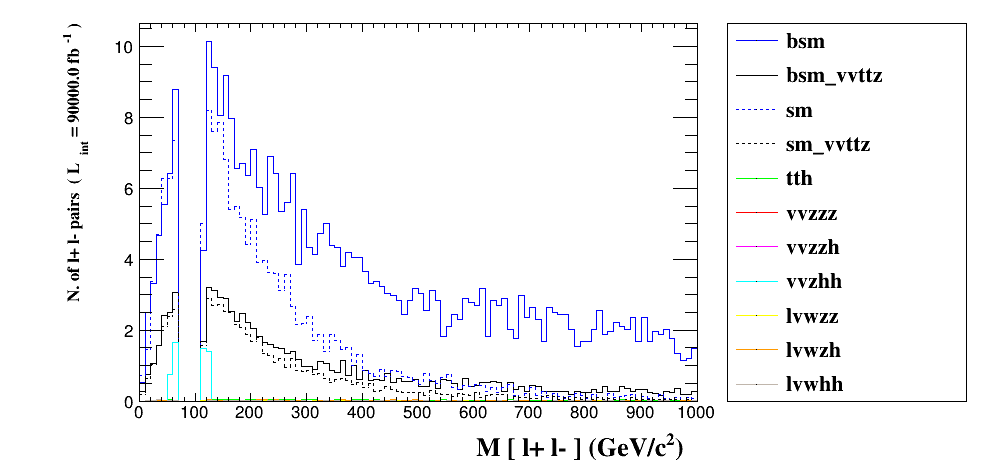}
    \includegraphics[width=80mm]{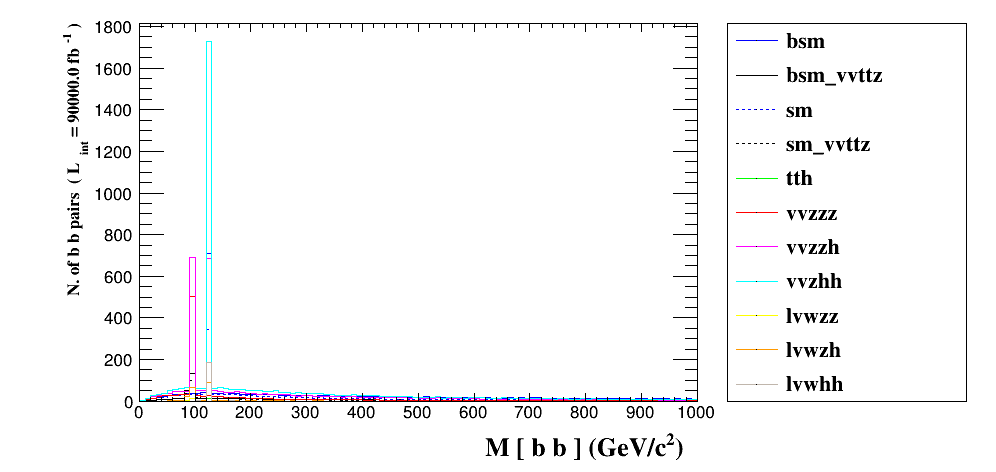}
    \includegraphics[width=80mm]{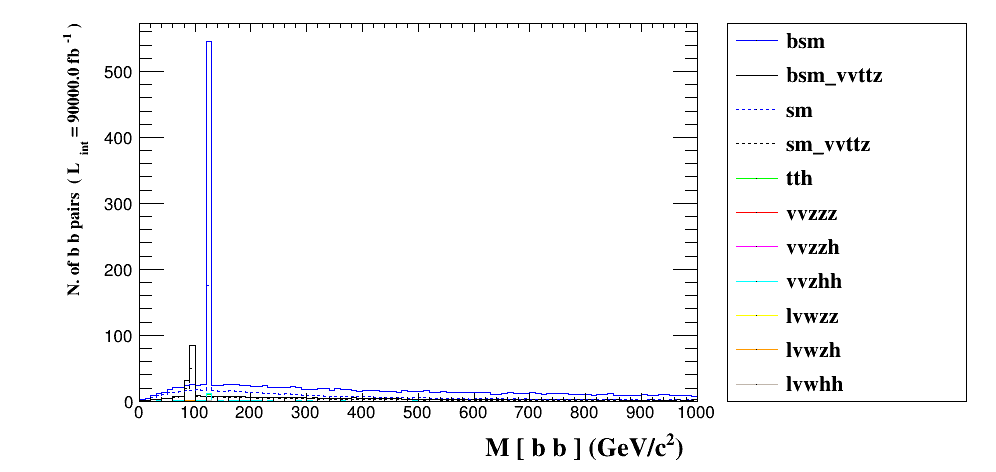}
    \caption{Distributions of the lepton channel of $\nu\nu t\bar th$ at 30 TeV before(left) and after(right) cuts (see Item(\ref{item:cuts_vvtth_lep_30})), see also Table(\ref{tab:cutflow_vvtth_lep_30}) for event numbers with the cut flow. }
    \label{fig:dis_vvtth_lep_30}
\end{figure}

To reduce background and increase statistical significance, we apply some cuts on the events, which are summarized as following:
\begin{itemize}\label{item:cuts_vvtth_lep_30}
    \item  Cut 1: Reject $70\  \text{GeV} <  M_{l^+ l^-} <  115 $ GeV.  The purpose of this cut is to reduce event numbers of   processes such as  $\nu\bar\nu Z_LZ_LZ_L$ that have $Z$ decaying into  $l^+ l^- $. 
    \item  Cut 2: Reject $\eta(l^+) >   3$; select $M_{b l^+} < 250 \ \ \text{GeV}$, select  $M_{b l^-} < 250$ GeV. The purpose of this cut is to reduce event numbers of  processes such as $l^+\nu_lW^+Z_LZ_L$ that have leptons with large rapidity and have no top quark as an intermediate state. 
    \item  Cut 3: Select $p_T(b) > 50 $ GeV, select $p_T(l) > 50 $ GeV, select MET $>$ 30 GeV. The purpose of this cut is to reduce the SM part of signal processes $\nu\bar\nu t_R\bar t_R h/Z_L$ and $\nu\bar\nu t_L\bar t_L h/Z_L$.
\end{itemize}
with $b = b /\bar{b}$ and  $l = l^+ /l^-$. The results of the cut flow are summarized in Table(\ref{tab:cutflow_vvtth_lep_30}).  In Fig.(\ref{fig:dis_vvtth_lep_30}), we also show some plots such as $p_T(l^+)$ before and after cuts to demonstrate the effects of our method. 

After obtaining the final results after cuts,  combined with b tagging efficiency $\epsilon_b=0.9$ and spin tagging efficiencies $\epsilon_s$,  we can obtain the corresponding statistical significance.  Taking $\epsilon_s=0.9$ (for all particles with helicities selected) as an example, following Eq.(\ref{eq:signficance}), we obtain $\mathcal S =22.8(\sigma)$ if combining  events of $\nu\nu t\bar th$ and $\nu\nu t\bar tZ$. This is an encouraging result. Following the same procedure we can obtain the constraint on $c_{t\phi}$($\Lambda = 1$ TeV) with $\epsilon_s=0.9$ as     $ -0.56  \leq  c_{t\phi}\leq 0.53$ at $2\sigma$ and $ -0.43  \leq c_{t\phi}  \leq 0.5$ at $1\sigma$. Converting them into constraints on $\delta y_t$(Eq.(\ref{eq:delyt_ctp})), we get 
$ -3.2\%\leq  \delta y_t\leq 3.4\%$ at $2\sigma$ and  $-3.1\%\leq \delta y_t \leq 2.6\%$ at  $1\sigma$. The results are quite promising.

\begin{table}[]
    \centering
    \begin{tabular}{|c|c|c|c|c|}
    \hline
    \hline
     process                & before cuts&  cut1 & cut2    & cut3 \\
     \hline
     BSM($\nu\bar \nu t_R\bar t_Rh/\nu\bar \nu t_L\bar t_Lh$) &  675.9$\times$2  & 610.8$\times$2 & 554.3$\times$2 & 524.3$\times$2 \\
     \hline
     BSM($\nu\bar \nu t_R\bar t_RZ_L/\nu\bar \nu t_L\bar t_LZ_L$) &  167.3$\times$2 & 143.8$\times$2 & 119.2$\times$2 & 107.9$\times$2  \\
     \hline
     \hline
     SM($\nu\bar \nu t_R\bar t_Rh/\nu\bar \nu t_L\bar t_L h$)  &  312.6$\times$2 & 248.8$\times$2 & 189.9$\times$2 & 159.1$\times$2  \\
    \hline 
    SM($\nu\bar \nu t_R\bar t_RZ_L/\nu\bar \nu t_L\bar t_LZ_L$)   &  117.4$\times$2 &  94.8$\times$2 &  71.3$\times$2 & 60.3$\times$2  \\      
    \hline
    $t_R\bar t_R h/t_L\bar t_L h$           &    10.5$\times$2    &  10.5$\times$2    &  10.5$\times$2  &  10.5$\times$2  \\
    \hline
    $\nu\bar\nu Z_LZ_LZ_L$          &    237.8          &    0      &     0      &     0                 \\
    \hline
    $\nu\bar\nu Z_LZ_Lh$           &     639.6         &     0     &       0   &       0      \\
    \hline
    $\nu\bar\nu Z_Lhh$          &   832.5    &   12.15   &   6.9    &   5.3       \\
    \hline
    $l^\pm\nu W^\mp Z_LZ_L$            &     8.0 $\times$ 2&     7.8 $\times$ 2&   0.7$\times$ 2&   0.7$\times$ 2\\
    \hline
    $l^{\pm}\nu W^{\mp}Z_Lh$      &    89.3$\times$2    &    88.6$\times$2     &  1.3$\times$2  &   1.3$\times$2   \\
    \hline 
    $l^{\pm}\nu W^{\mp}hh$     &   91.874$\times$2   &   91.3$\times$2     &    0.7$\times$2 &   0.7$\times$2   \\
    \hline
    \end{tabular}
    \caption{Cut flow of the lepton channel of $\nu\nu t\bar th$ at 30 TeV with integrated luminosity $90 \ ab^{-1}$.  BSM processes are set at $c_{t\phi} = 2$ (with $\Lambda =1$ TeV).  Helicity selections are shown in the table. The b tagging efficiency and spin tagging efficiencies $\epsilon_s$ are set to be is $\epsilon_b=\epsilon_s=1$. }
    \label{tab:cutflow_vvtth_lep_30}
\end{table}

\begin{figure}
    \centering
    \includegraphics[width=170mm]{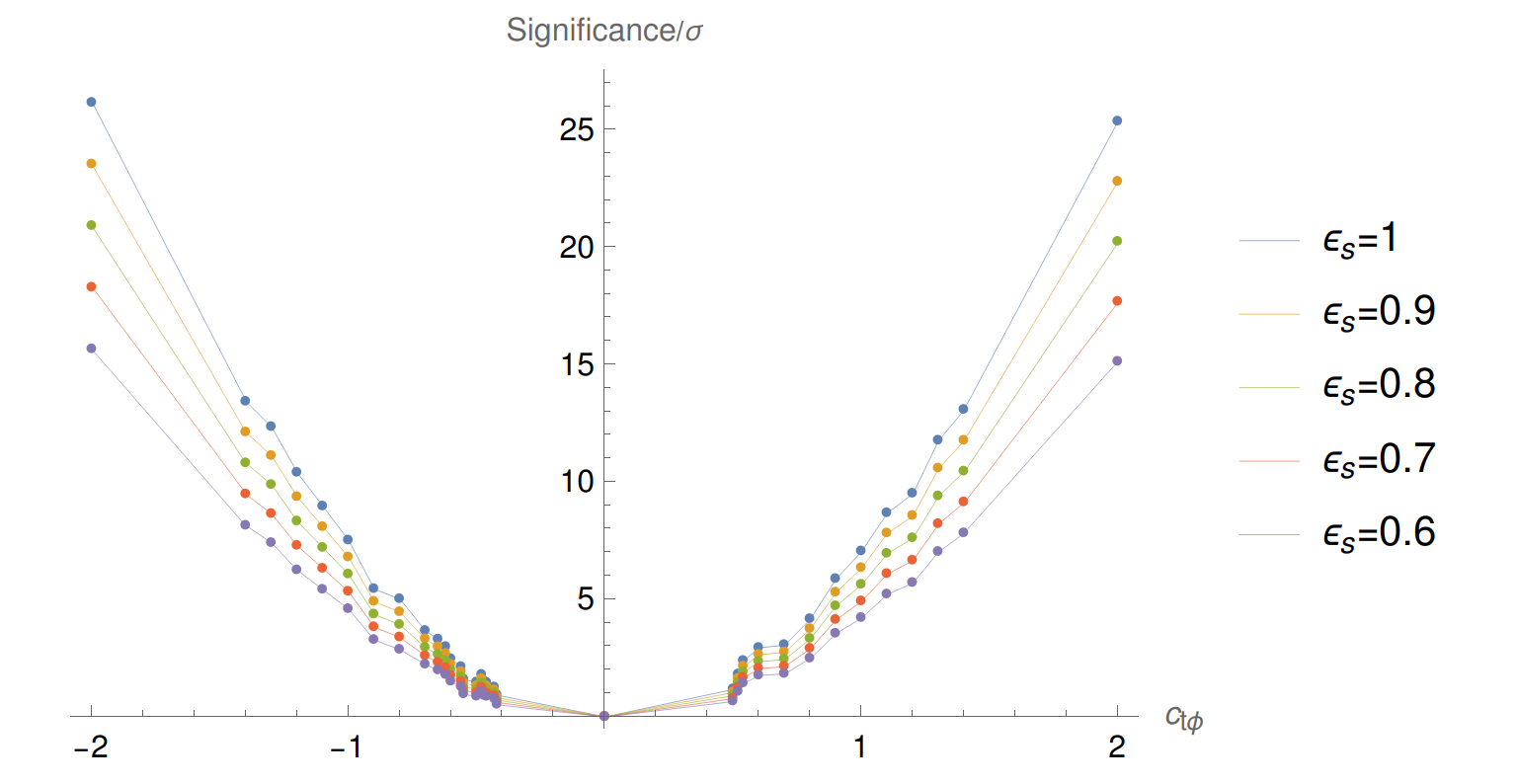}
    \caption{The dependence of statistical significance of the lepton channel of $\nu\nu t\bar th$ at 30 TeV(Item(\ref{item:cuts_vvtth_lep_30})) on  $c_{t\phi}$ ($\Lambda = 1$ TeV) and spin tagging efficiency $\epsilon_s$.  $Z$ boson and $t/\bar t$ take the same value on $\epsilon_s$. The b quark  tagging efficiency is set to be $\epsilon_b=0.9$. }
    \label{fig:sig_vvtth_lep_30}
\end{figure}

Finally, the technique for tagging the spins of a particle is not yet mature and subject to variation for different processes, thus we analyze how statistical significance $\mathcal S$ changes with spin tagging efficiency $\epsilon_s$, as well as with  $c_{t\phi}$. For the values of $\epsilon_s$, we set equal for that of  $Z$ and $t/\bar t$ and scan over $1, 0.9, 0.8, 0.7, 0.6$. The results are summarized in Fig(\ref{fig:sig_vvtth_lep_30}). The results show that the statistical significance remains high when spin tagging efficiency varies, even when reaching as low as $\epsilon_s=0.6$. 
We also compute the limits of $\delta y_t$ at $\epsilon_s=0.7$, the results are summarized with $\epsilon_s=0.9$ and other processes in Table(\ref{tab:constraints_summary}).

\newpage
\subsubsection{10 TeV}
We now turn to 10 TeV for the lepton channel from $\nu\nu t\bar th$. We impose the following cuts to reduce background:
\begin{itemize}\label{item:cuts_vvtth_lep_10}
  \item  Cut 1: Reject $70\  \text{GeV} <  M_{l^+ l^-} <  115 $ GeV, reject MET $>$ 650 GeV
  \item  Cut 2: Select $p_T(b) > 50 $ GeV, select $p_T(l) > 50 $ GeV.
\end{itemize}
with  $b = b /\bar{b}$ and  $l = l^+ /l^-$. 

The cuts on $10$ TeV is modified slightly relative to 30 TeV of the same channel: Cut 1 is to reduce background processes, while Cut 2 is to optimize signals relative to their SM counter parts. The cut flow of the signal and background processes at $c_{t\phi}=2$($\Lambda=1$ TeV) is summarized in Table(\ref{tab:cutflow_vvtth_lep_10}).  Using the cuts in Item(\ref{item:cuts_vvtth_lep_10}) and setting the spin tagging efficiency to be $0.9$, $\delta y_t$  is  constrained to be within $ [-11.0\%, 9.8\% ]$ at $1\sigma$ level.  At $2\sigma$ level,  didn't take  the precise value, since we have $|c_{t\phi}|>2$ ($|\delta_{y_t}| > 12.2 \%$). 

\begin{table}[b]
    \centering
    \begin{tabular}{|c|c|c|c|}
    \hline
    \hline
     process                & before cuts&  cut1 & cut2   \\
     \hline
     BSM($\nu\bar \nu t_R\bar t_Rh/\nu\bar \nu t_L\bar t_Lh$) &  15.3$\times$2  & 9$\times$2  & 8.1$\times$2   \\
     \hline
     BSM($\nu\bar \nu t_R\bar t_RZ_L/\nu\bar \nu t_L\bar t_LZ_L$) &  5.1$\times$2 & 2.6$\times$2 & 2.2$\times$2  \\
     \hline
     \hline
     SM($\nu\bar \nu t_R\bar t_Rh/\nu\bar \nu t_L\bar t_L$)  &  11.2$\times$2 & 5.6$\times$2 & 4.7$\times$2   \\
    \hline 
    SM($\nu\bar \nu t_R\bar t_RZ_L/\nu\bar \nu t_L\bar t_LZ_L$)   &  4.6$\times$2 &  2.2$\times$2 &  1.8$\times$2   \\      
    \hline
    $t_R\bar t_R h/t_L\bar t_L h$           &   7.4$\times$2  & 2.6$\times$2  & 2.6$\times$2   \\
    \hline
    $t_R\bar t_R Z_L/t_L\bar t_L Z_L$      &1.3$\times$2     &   0.6$\times$2    &  0.5$\times$2    \\
    \hline
    $\nu\bar\nu Z_LZ_LZ_L$          &    8.6          &    0      &     0       \\
    \hline
    $\nu\bar\nu Z_LZ_Lh$           &     22.9        &     0     &       0   \\
    \hline
    $\nu\bar\nu Z_Lhh$          &   31.2    &   0.3   &   0.2         \\
    \hline
    $lvwzz$                    &  1.7       &    0.3    &      0.3        \\
    \hline 
    $lvwzh$                   &   5.0      &     1.3   &       1.3        \\
    \hline 
    $lvwhh$                    &   6.1     &     2.0   &       2.0       \\
    \hline
    \end{tabular}
    \caption{Cut flow of lepton channel of $\nu\nu t\bar th$ at 10 TeV with integrated luminosity $10 \ ab^{-1}$. 
BSM processes are simulated at $c_{t\phi}=2$ (with $\Lambda = 1$ TeV). Helicity selections are shown in the table. Processes with negligible cross sections are not listed. The b tagging efficiency and spin tagging efficiencies $\epsilon_s$ are set to be is $\epsilon_b=\epsilon_s=1$.
}
    \label{tab:cutflow_vvtth_lep_10}
\end{table}

\newpage
\subsection{Semi-Lepton Channel}

In this subsection, we study  the channel $l^\pm jj4b+\text{MET}$ with $l=e,\mu$, which has contributions from both  $\nu\nu t\bar th$ and $\nu\nu t\bar tZ$, but  $\nu\nu t\bar th$  is the dominant process. It's obtained through $t\rightarrow b W^+\rightarrow b l^+ \nu_l$, $\bar{t}\rightarrow \bar{b} W^-\rightarrow \bar b jj$, $h/Z_L\rightarrow b \bar{b}$; or $t\leftrightarrow \bar t $. As usual, we choose the helicities of $t,\bar t$ and $Z$ boson to be $t_R\bar t_R/t_L\bar t_L$ and $Z_L$ respectively, while the helicities of $W^\pm$ are not tagged.  The spin tagging efficiencies of all particles are set to be the same. 

The main background processes are
\begin{eqnarray}
    &&\mu\mu\rightarrow \nu\bar \nu t_R\bar{t}_Rh/Z_L(\text{SM}) \rightarrow  3\nu l^{\pm} 4bjj\\
    &&\mu\mu\rightarrow t_R\bar t_R h/Z_L \rightarrow  2\nu l^{\pm} 4bjj\\
    &&\mu\mu\rightarrow l^+\nu_lW^-Z_LZ_L(/l^-\bar\nu_l W^+Z_LZ_L) \rightarrow \nu l^{\pm}4bjj\\
    &&\mu\mu\rightarrow l^+\nu_lW^-Z_Lh(/l^-\bar\nu_l W^+Z_Lh) \rightarrow \nu l^{\pm}4bjj\\
    &&\mu\mu\rightarrow l^+\nu_lW^-hh(/l^-\bar\nu_l W^+
    hh) \rightarrow \nu l^{\pm}4bjj
\end{eqnarray}


\subsubsection{30 TeV}

We start with  c.m. energy of 30 TeV.   The semi-lepton channel has  a much larger cross-section than lepton channel due to the larger branching ratio of $W\rightarrow jj$. With the increase of luminosity and smaller number of background processes, we expect a  high  significance and stringent limit on $\delta y_t$.   

\begin{table}[]
    \centering
    \begin{tabular}{|c|l|c|c|l|}
    \hline
    \hline
     process                &  before cuts&cut1& cut2     &     cut3\\
     \hline
     BSM($\nu\bar{\nu} t_R\bar t_Rh/\nu\bar \nu t_L\bar t_L h$) &   2025$\times$4&1963.41$\times$4& 1695.2$\times$4&1239.2 $\times$4\\
     \hline
     BSM($\nu\bar{\nu} t_R\bar t_RZ_L/\nu\bar \nu t_L\bar t_L Z_L$)&   505$\times$4&481.30$\times$4& 405.14 $\times$4&257.0 $\times$4\\
     \hline
     \hline
     SM($\nu\bar{\nu} t_R\bar t_Rh/\nu\bar \nu t_L\bar t_L h$)  &   884$\times$4&829.1$\times$4& 706.5 $\times$4&333.5 $\times$4\\
    \hline 
    SM($\nu\bar{\nu} t_R\bar t_RZ_L/\nu\bar \nu t_L\bar t_L Z_L$)&   352$\times$4&329.55$\times$4&  273.86 $\times$4&135.32 $\times$4\\      
    \hline
    $t_R\bar{t}_R h/t_L\bar t_L h$           &     32.2$\times$4&32.181$\times$4&  17.49 $\times$4&16.36 $\times$4\\
    \hline
    $t_R\bar{t}_R Z_L/t_L\bar t_L Z_L$&     5.66$\times$4&5.6557 $\times$4&     2.64 $\times$4&2.48 $\times$4\\
    \hline
    $l^\pm\nu W^\mp Z_LZ_L$&     22.3$\times$2&22.259 $\times$2&   4.39 $\times$2&3.13 $\times$2\\
    \hline
    $l^\pm \nu W^\mp Z_Lh$&     273$\times$2&272.42 $\times$2&  61.77 $\times$2&55.85 $\times$2\\
    \hline 
    $l^\pm\nu W^\mp hh$   &    304$\times$2&304.15 $\times$2&    171.99 $\times$2&168.9 $\times$2\\
    \hline
    \end{tabular}
     \caption{Cut flow of the semi-lepton channel of $\nu\nu t\bar th$ at 30 TeV, with integrated luminosity $90 \ ab^{-1}$. 
BSM processes are simulated at $c_{t\phi}=2$ (with $\Lambda = 1$ TeV). Helicity selections are shown in the table. The b tagging efficiency and spin tagging efficiencies $\epsilon_s$ are set to be is $\epsilon_b=\epsilon_s=1$.
       }
    \label{tab:cutflow_vvtth_semi_30}
    \end{table}

\begin{figure}
    \centering
    \includegraphics[width=80mm]{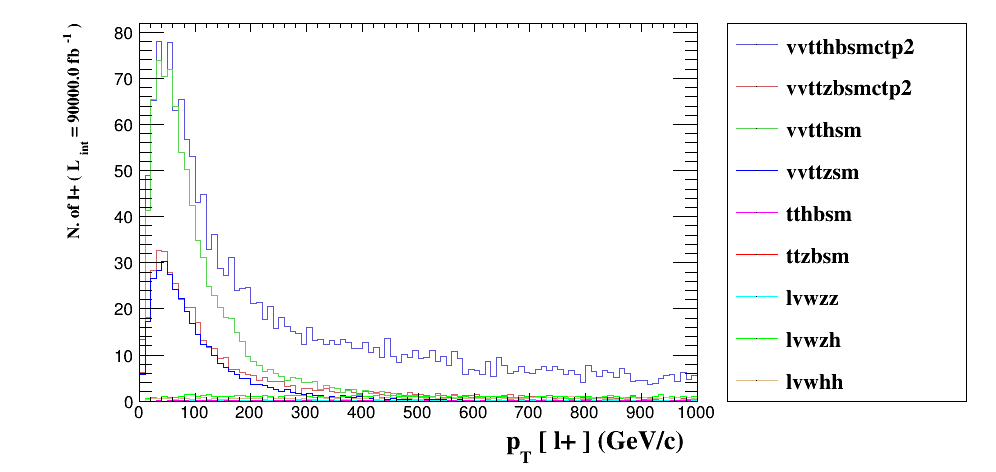}
    \includegraphics[width=80mm]{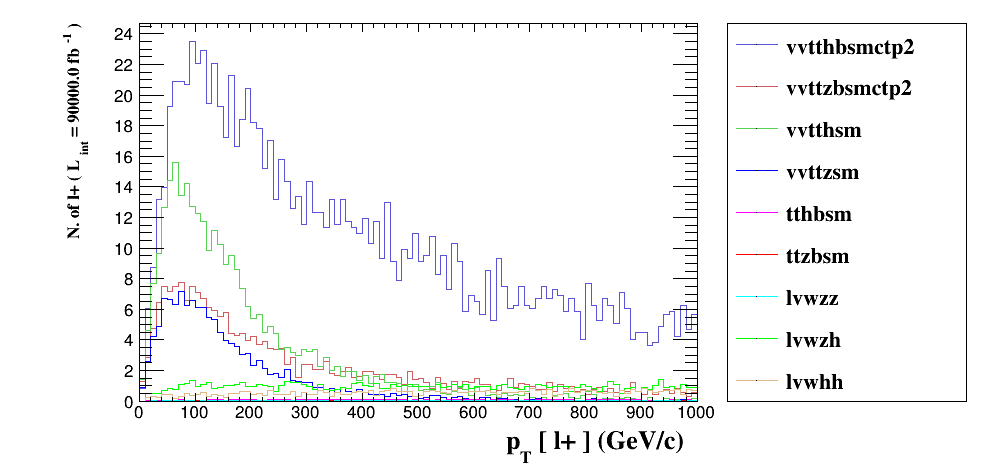}
    \includegraphics[width=80mm]{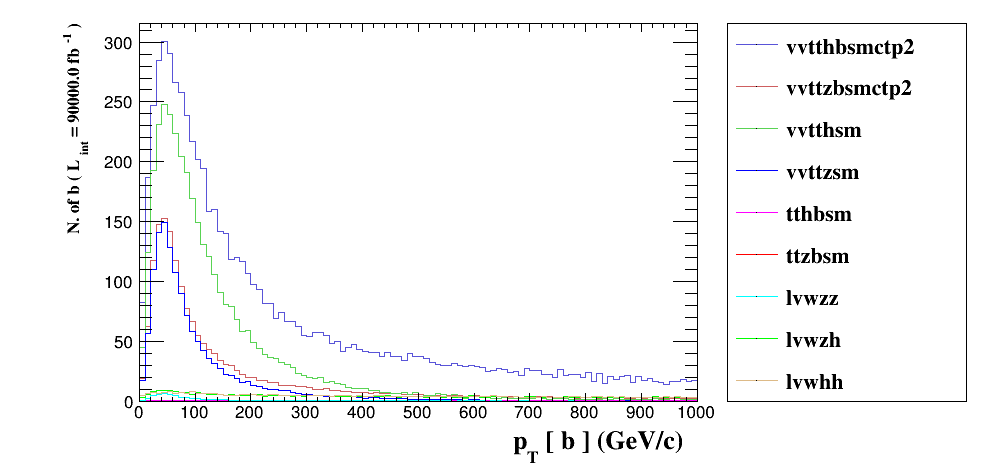}
    \includegraphics[width=80mm]{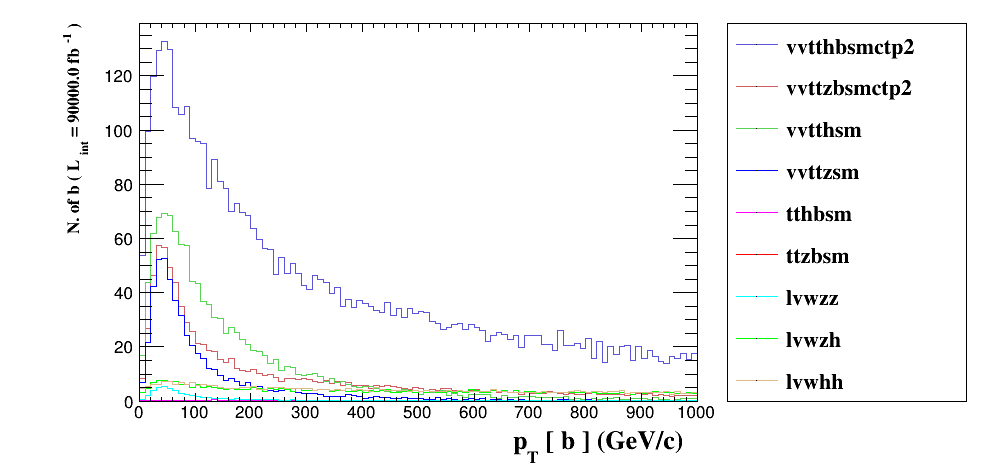}
    \includegraphics[width=80mm]{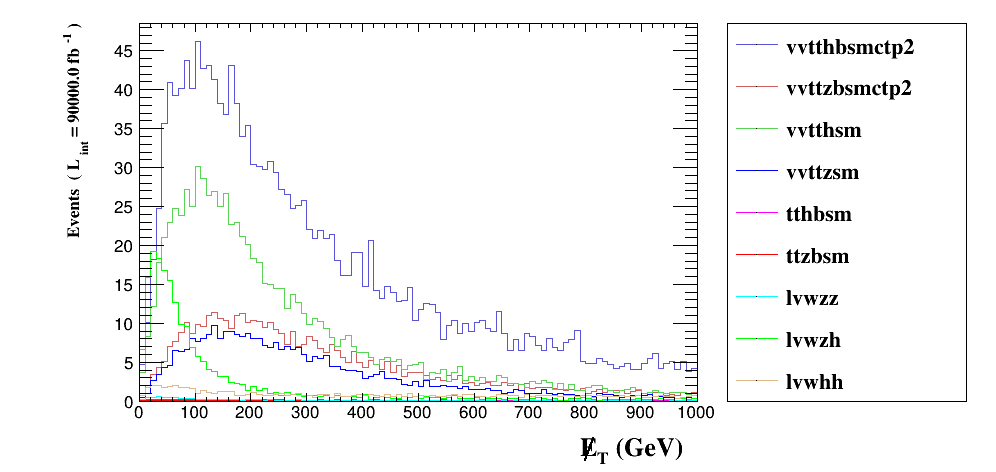}
    \includegraphics[width=80mm]{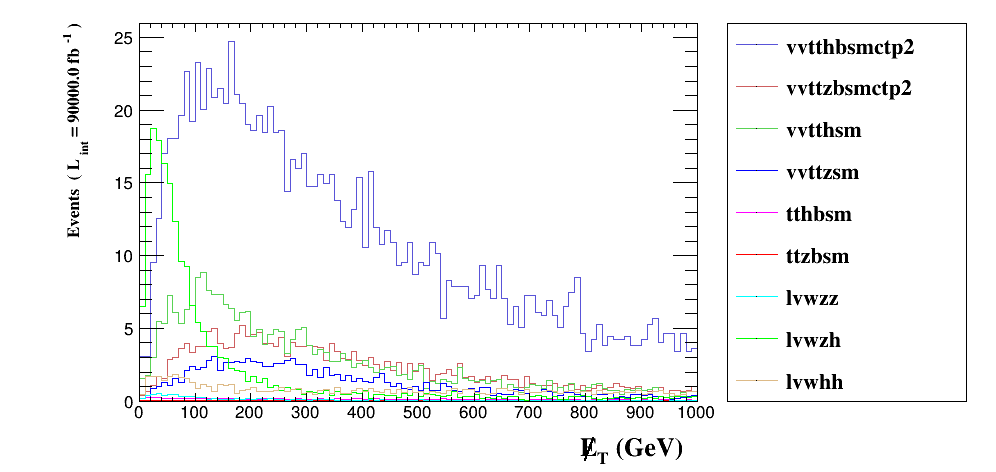}
    \includegraphics[width=80mm]{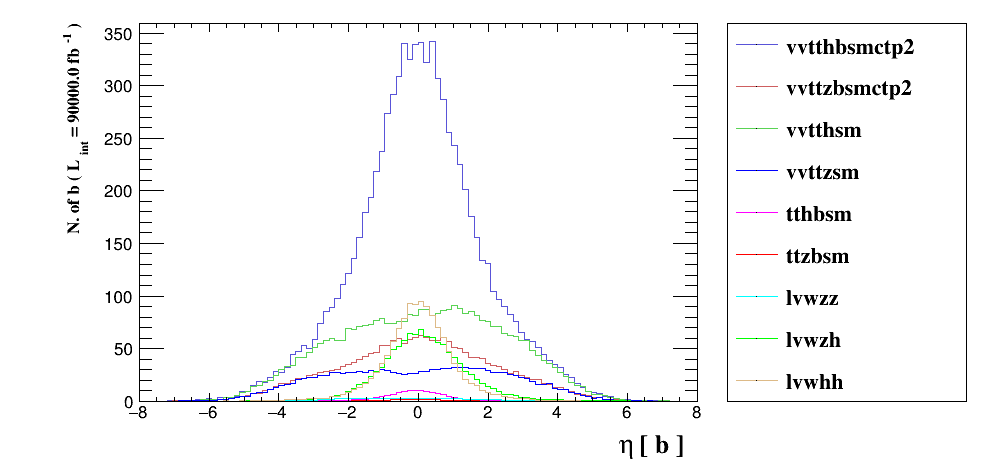}
    \includegraphics[width=80mm]{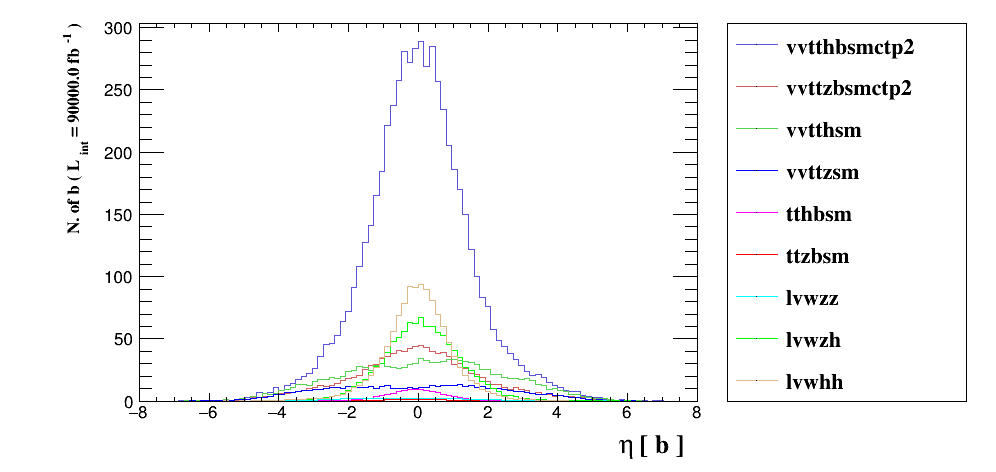}
     \includegraphics[width=80mm]{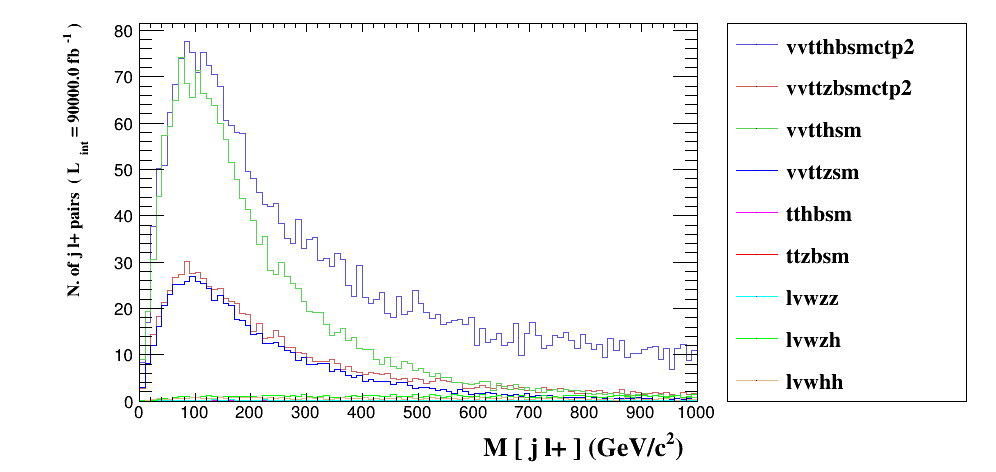}
     \includegraphics[width=80mm]{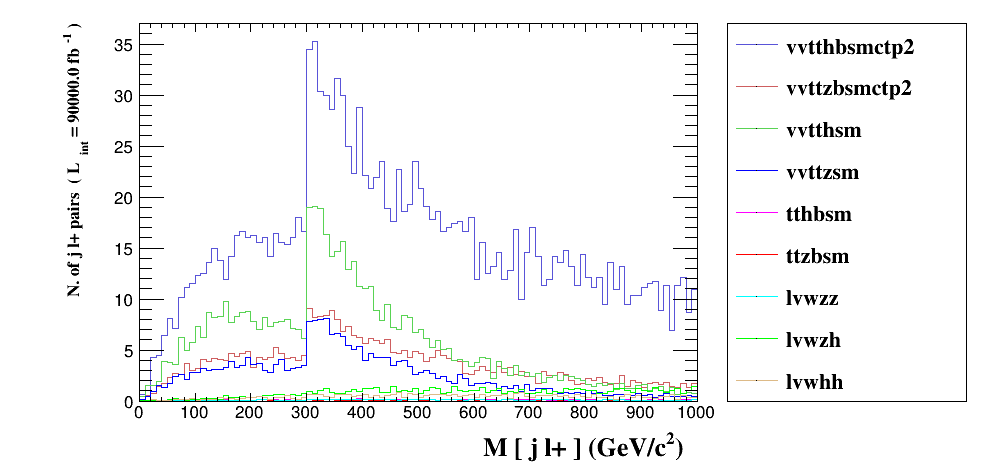}
    \caption{Distributions of the semi-lepton channel $\nu\nu t\bar th$ at 30 TeV before(left) and after(right) cuts (Item(\ref{item:cuts_vvtth_semi_30})), see also Table(\ref{tab:cutflow_vvtth_semi_30}) for event numbers with the cut flow. }
    \label{fig:dist_vvtth_semi_30}
\end{figure}

 As summarized as below, we apply the following cuts to reduce background and increase statistical significance:
   \begin{itemize}\label{item:cuts_vvtth_semi_30}
    \item  Cut 1: reject $p_T(l) < 20$ GeV.  This cut reduces both SM and BSM events of $\nu\nu t\bar th/Z$ equally, therefore enhancing the signal relative to the background. 
    \item  Cut 2: select $p_T(b) <$ 700 GeV, reject 1.7 $< \eta(l^+) <$ 2.5. This cut reduces $l\nu wzh$ and $l\nu whh$, as well as the SM part of $\nu\nu t\bar th$ and $\nu\nu t\bar tz$.  
    \item Cut 3: select -2 $< \eta(b) <$ 2, reject $M(b j j ) < 200$ GeV, reject MET $<$ 20 GeV. In this cut we continue to optimize the results by  reducing SM of $\nu\nu t\bar th/Z$.  
\end{itemize}
with $b = b /\bar{b}$ and  $l = l^+ /l^-$.  The results of the cut flow are summarized in Table(\ref{tab:cutflow_vvtth_semi_30}),  with which we can then obtain the corresponding statistical significance. Taking $\epsilon_s=0.9$ (and $\epsilon_b=0.9$), following Eq.(\ref{eq:signficance}), we obtain $\mathcal S=50.27(\sigma)$ at $c_{t\phi}=2$ and $\Lambda = 1$ TeV if combining $\nu\nu t\bar th$ and $\nu\nu t\bar tZ$. We also show some plots of the events before and after the cuts in Fig.(\ref{fig:dist_vvtth_semi_30}).  Following the same procedure we can obtain the constraint on $C_{t\phi}\equiv \frac{c_{t\phi}}{\Lambda^2}$ with $\epsilon_s=0.9$ as  $ -0.61\  \text{TeV}^{-2}\leq  C_{t\phi}\leq 0.625 \ \text{TeV}^{-2}$ at $5\sigma$,   $ -0.45\  \text{TeV}^{-2}\leq  C_{t\phi}\leq 0.405\  \text{TeV}^{-2}$ at $2\sigma$ and $ -0.3\  \text{TeV}^{-2}\leq  C_{t\phi}\leq 0.27\  \text{TeV}^{-2}$ at $1\sigma$. Converting them into constraints on $\delta y_t$ using Eq.(\ref{eq:delyt_ctp}), we get $ -3.8\% \leq  \delta y_t\leq 3.7\% $ at $5\sigma$, $ -2.5\%\leq  \delta y_t\leq 2.7\%$ at $2\sigma$ and  $ -1.6\%\leq  \delta y_t\leq 1.8\%$ at $1\sigma$. As we can see, the results are visibly better than  the lepton channel.

\begin{figure}
    \centering
    \includegraphics{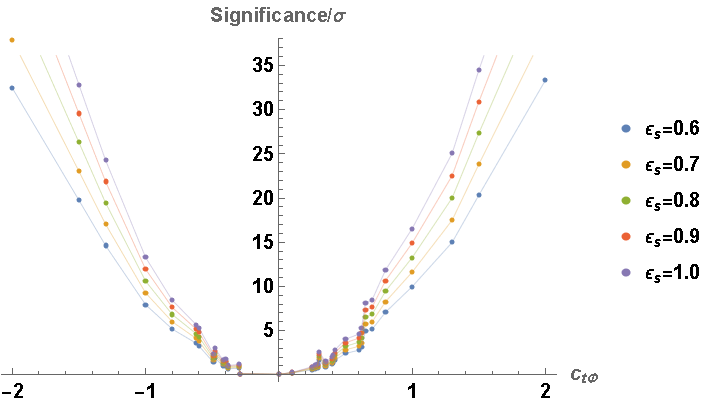}
    \caption{The dependence of statistical significance of the semi-lepton channel $\nu\nu t\bar th$ at 30 TeV on  $c_{t\phi}$ (with $\Lambda = 1$ TeV) and spin tagging efficiency $\epsilon_s$, the latter of $Z$ and $t/\bar t$  are take to be the equal. The b quark  tagging efficiency is set to be $\epsilon_b=0.9$. }
    \label{fig:sig_vvtth_semi_30}
\end{figure}

Finally,  we analyze how statistical significance $\mathcal S$ changes with spin tagging efficiency $\epsilon_s$, as well as with  $c_{t\phi}$.  
The results are summarized in Fig(\ref{fig:sig_vvtth_semi_30}).

\newpage
\subsubsection{10 TeV}
\begin{figure}
    \centering
    \includegraphics[width=80mm]{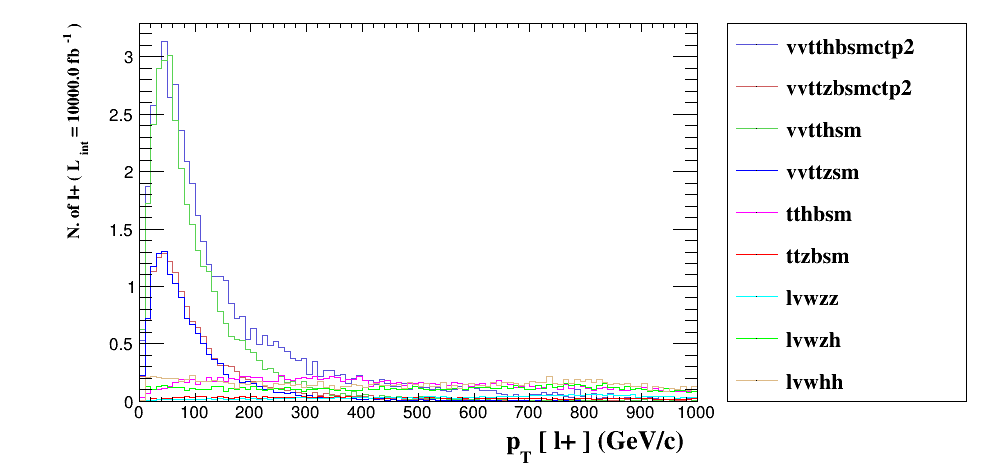}
    \includegraphics[width=80mm]{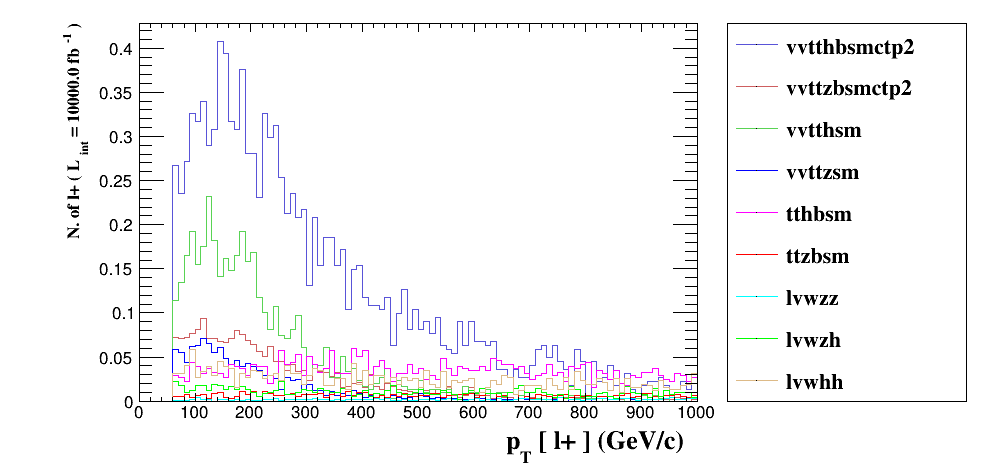}
     \includegraphics[width=80mm]{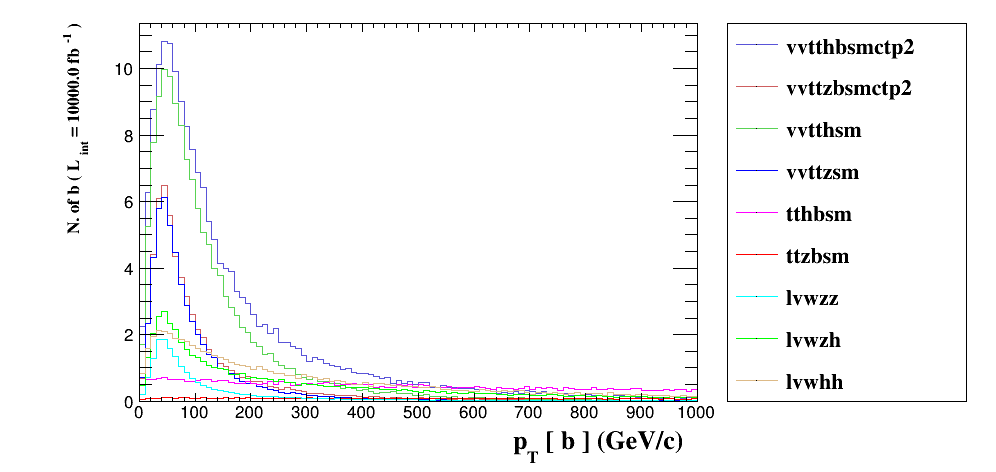}
      \includegraphics[width=80mm]{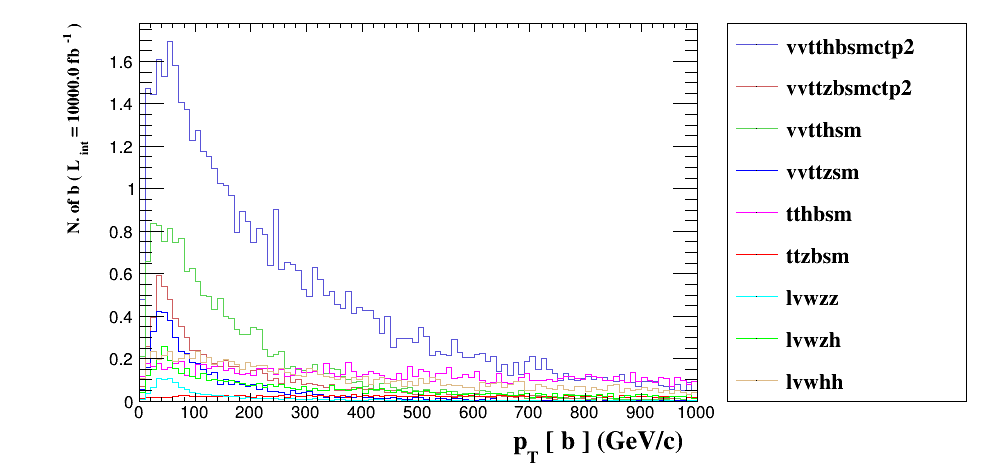}
    \includegraphics[width=80mm]{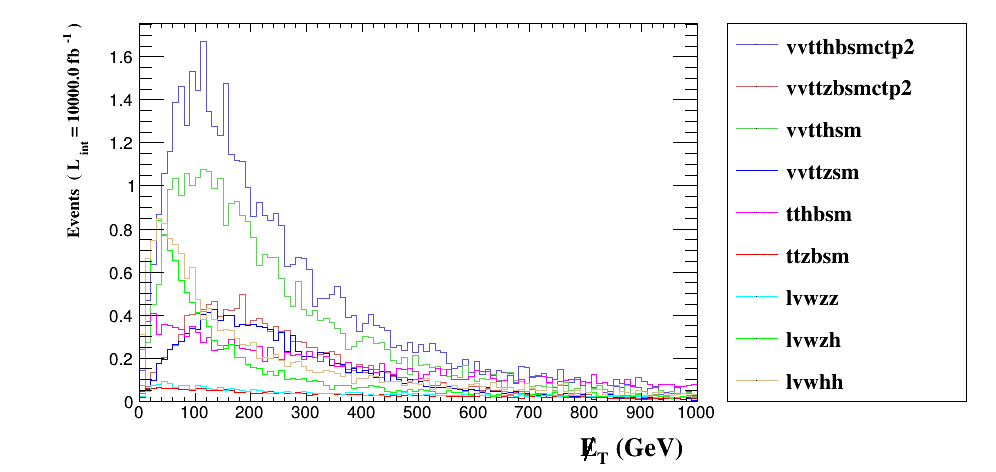}
    \includegraphics[width=80mm]{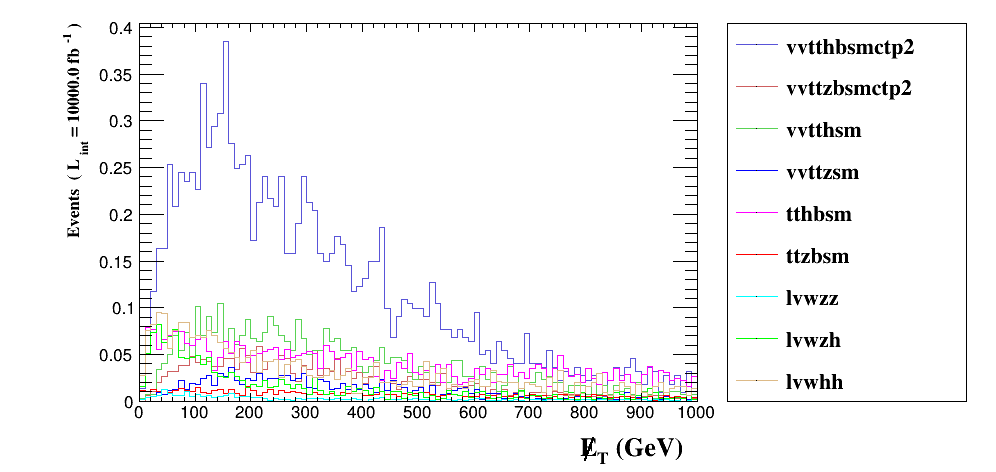}
    \includegraphics[width=80mm]{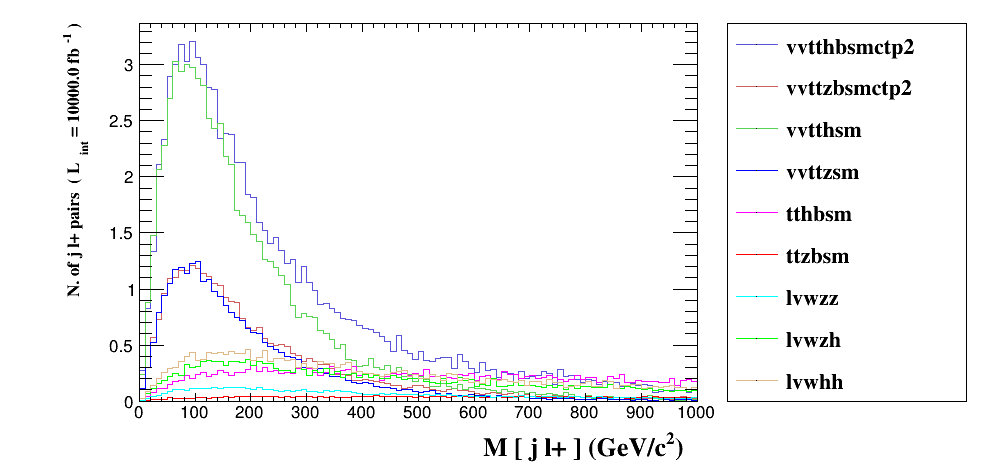}
    \includegraphics[width=80mm]{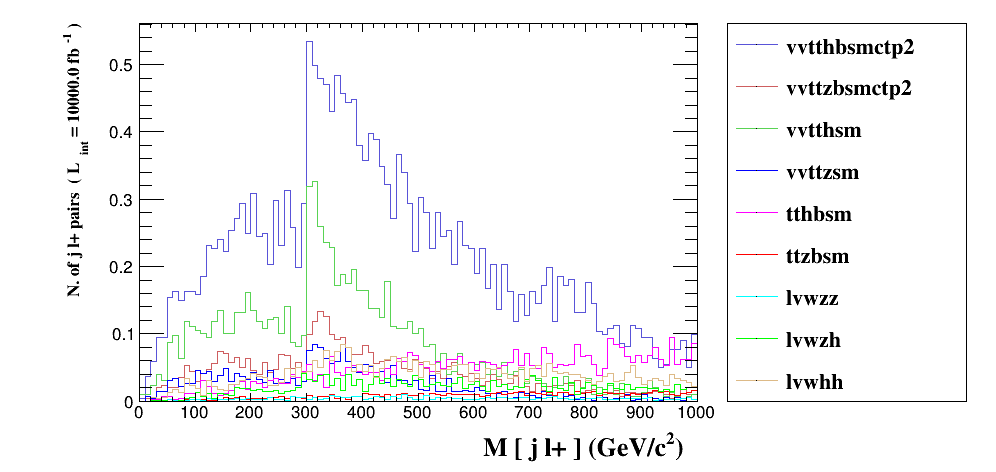}
    \includegraphics[width=80mm]{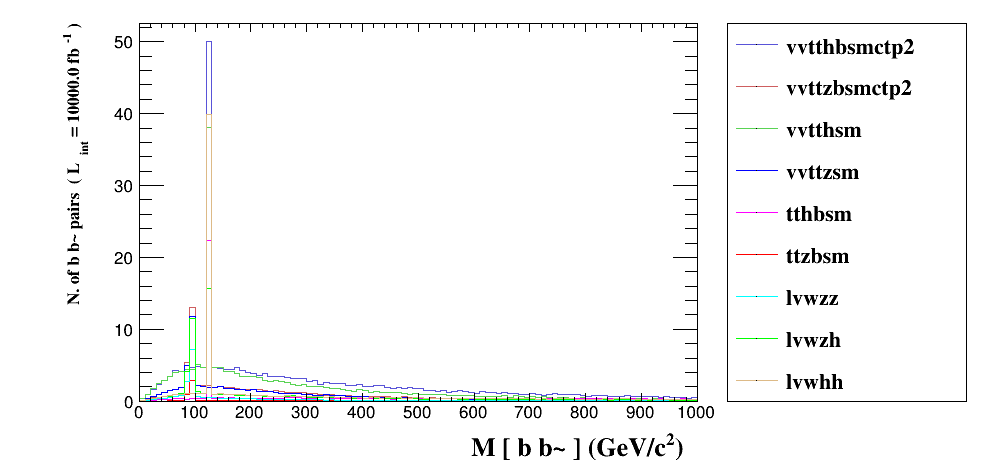}
    \includegraphics[width=80mm]{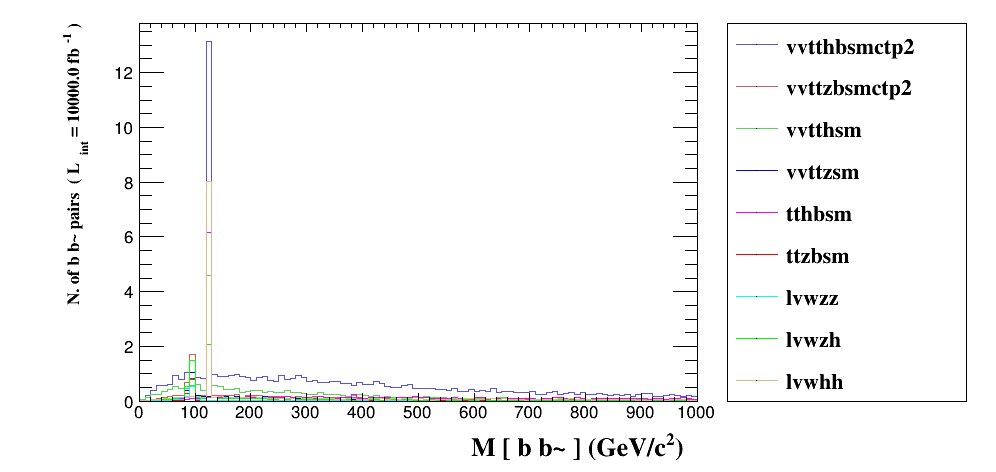}
    \caption{Distributions of the semi-lepton channel from $\nu\nu t\bar th$ at 10 TeV before(left) and after(right) cuts (Item(\ref{item:cuts_vvtth_semi_10})), see also Table(\ref{tab:cutflow_vvtth_semi_10}) for event numbers with the cut flow. The b tagging efficiency and spin tagging efficiency are set to be $\epsilon_b=\epsilon_s=1$.}
    \label{fig:dist_vvtth_semi_10}
\end{figure}

We then proceed to $10$ TeV for the semi-lepton channel of $\nu\nu t\bar th$. 
The cuts we applied to reduce background and increase statistical significance are summarized as follows:

\begin{itemize}\label{item:cuts_vvtth_semi_10}
    \item  Cut 1: Select $-2 < \eta(b) <   2$, Select $ M_{l^+ l^-} > 300\  \text{GeV} $, Select $ M_{b j} > 200\  \text{GeV}$. The purpose of this cut is to reduce the SM part of signal processes $\nu\bar\nu t_R\bar t_R h/Z_L$ and $\nu\bar\nu t_L\bar t_L h/Z_L$.
  \item Cut 2: Select $2 < \Delta R_{l^+ ,j} < 4$, $\Delta R_{l^+ ,b} > 4$, $1 < \Delta R_{b ,j} < 4$. The purpose of this cut is to reduce the contribution from processes such as $l^+\nu_lW^+Z_LZ_L$, $l^+\nu_lW^+Z_Lh$, $l^+\nu_lW^+hh$ which typically produce leptons with large rapidity and do not involve top quarks as intermediate states. This selection helps  isolating the signal processes of interest by removing events where the lepton-jet separation $\Delta R_{l^+ ,j}$ is either too small or too large, and ensuring that the lepton-b quark separation $\Delta R_{l^+ ,b}$ is sufficiently large to exclude non-top quark background. Additionally, requiring $1 < \Delta R_{b ,j} < 4$ helps  suppressing background from SM processes and $t\bar{t}h$ production.
    \item  Cut 3: Select $p_T(l^+) > 60 \  \text{GeV} $,  Select $800 \ \text{GeV} < p_T(j) < 1000 \ \text{GeV}$, Select  $\eta(l) > 2$. Without rejecting a large number of signal events, it make some cuts and optimize the significance.
\end{itemize}
with $b = b /\bar{b}$ and  $l = l^+ /l^-$. The results of the cut flow are summarized in Table(\ref{tab:cutflow_vvtth_semi_10}), according to which  we can obtain the corresponding statistical significance. The b tagging efficiency and spin tagging efficiency are set to be $\epsilon_b=\epsilon_s=0.9$ (for all particles with helicities specified) and following Eq.(\ref{eq:signficance}), we obtain $\mathcal S=3.2(\sigma)$ at $C_{t\phi} = 2 \ \text{TeV}^{-1}$.  Following the same procedure as the previous sections, we can obtain the constraint on $c_{t\phi}$ with $\epsilon_s=0.9$ as   $ -1.6 \leq  c_{t\phi}\leq 1.6 $ at $2\sigma$ and $ -1.1 \leq  c_{t\phi}\leq 1.15 $ at $1\sigma$ with $\Lambda =1$ TeV. Converting them into constraints on $\delta y_t$,  we get $ -9.8\%\leq  \delta y_t\leq 9.8\%$ at $2\sigma$ and $ -7.0\%\leq  \delta y_t\leq 6.7\%$ at $1\sigma$. We also show some plots of the events before and after the cuts in Fig.(\ref{fig:dist_vvtth_semi_10}).

\begin{table}[]
    \centering
     \begin{tabular}{|c|c|c|c|l|}
    \hline
    \hline
     process                & before cuts&  cut1 & cut2     &            cut3\\
     \hline
     BSM($\nu\bar{\nu} t_R\bar t_Rh/\nu\bar \nu t_L\bar t_L h$) &  45.3$\times$4& 19.67$\times$4& 14.61$\times$4&12.2$\times$4\\
     \hline
     BSM($\nu\bar{\nu} t_R\bar t_RZ_L/\nu\bar \nu t_L\bar t_L Z_L$) &  15.4$\times$4& 5.41$\times$4& 2.79$\times$4&2.18$\times$4\\
     \hline
     \hline
     SM($\nu\bar{\nu} t_R\bar t_Rh/\nu\bar \nu t_L\bar t_L h$)  &  33.7$\times$4& 9.02$\times$4& 5.29 $\times$4&4.15$\times$4\\
    \hline 
    SM($\nu\bar{\nu} t_R\bar t_RZ_L/\nu\bar \nu t_L\bar t_L Z_L$)   &  13.7$\times$4&  3.91$\times$4&  1.61 $\times$4&1.23$\times$4\\      
    \hline
    $t_R\bar t_R h/t_L\bar t_R h$           &    22.1$\times$4&  20.5$\times$4&  13.59 $\times$4&6.07$\times$4\\
    \hline
    $t_R\bar t_R Z_L/t_L\bar t_L Z_L$&    3.93$\times$4&    3.727 $\times$4&     2.612 $\times$4&1.14$\times$4\\
    \hline
    $l\nu WZ_LZ_L$            &     4.62$\times$2&     3.499 $\times$2&   0.79 $\times$2&0.354$\times$2\\
    \hline
    $l\nu WZ_Lh$&    14.7$\times$2&    11.77 $\times$2&  4.14 $\times$2&1.97$\times$2\\
    \hline 
    $l\nu Whh$&   19.5$\times$2&   15.85 $\times$2&    7.48 $\times$2&3.96$\times$2\\
    \hline
    \end{tabular}
    \caption{Cut flow of the semi-lepton channel of $\nu\nu t\bar th$ at 10 TeV, with integrated luminosity $10 \ ab^{-1}$. 
    BSM processes are simulated at $c_{t\phi}=2$ (with $\Lambda = 1$ TeV). Helicity selections are shown in the table. The b tagging efficiency and spin tagging efficiency are set to be $\epsilon_b=\epsilon_s=1$. }
    \label{tab:cutflow_vvtth_semi_10}
\end{table}

\begin{figure}
    \centering
    \includegraphics{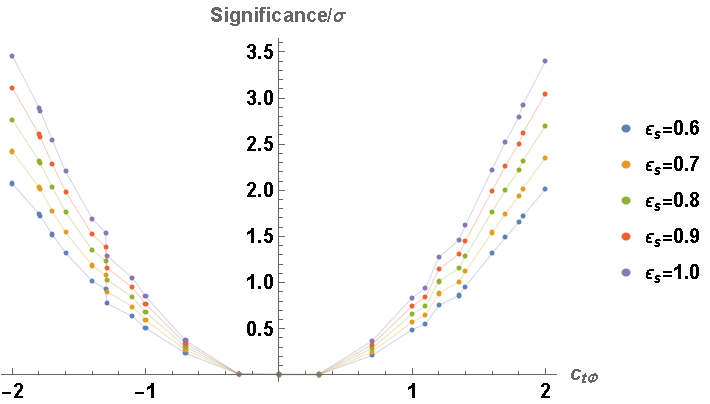}
    \caption{The dependence of statistical significance of the semi-lepton channel of $\nu\nu t\bar th$ at 10 TeV on  $c_{t\phi}$ (with $\Lambda = 1$ TeV)  and spin tagging efficiency $\epsilon_s$, the latter of $Z$ and $t/\bar t$  are take to be  equal. The b quark  tagging efficiency is set to be $\epsilon_b=0.9$. }
    \label{fig:sig_vvtth_semi_10}
\end{figure}

Finally, we analyze how statistical significance $\mathcal S$ changes with spin tagging efficiency $\epsilon_s$ and $c_{t\phi}$. For the values of $\epsilon_s$, we set equal for that of  $Z$ and $t/\bar t$ and scan over $1, 0.9, 0.8, 0.7, 0.6$. The results are summarized in Fig(\ref{fig:sig_vvtth_semi_10}).

\pagebreak

\section{Full Analysis of $\nu\nu t\bar tZ$}
\label{sec:vvttz}
In this section we analyze the lepton channel of $2(l^+l^-)2b+\text{MET}$ and the semi-lepton channel of $3l^{\pm}jj2b+\text{MET}$ with $l=e,\mu$, $b=b/\bar b$, which come from  $\nu\nu t\bar tZ$ and $\nu\nu t\bar th$, but dominated by the former.  Our simulation strategy and settings are the same as Sec.(\ref{sec:parton_signf} and \ref{sec:vvtth}), so we won't repeat here.

\subsection{Lepton Channel }

In this subsection, we study the channel $2(l^+l^-)2b+\text{MET}$, which comes from $\nu\nu t\bar tZ$ only through $t\rightarrow b W^+\overset{W^+ \rightarrow l^+\nu_l}{\longrightarrow} b l^+ \nu_l$, $\bar t\rightarrow \bar b W^-\overset{W^- \rightarrow l^-\bar \nu_l}{\longrightarrow} \bar b l^- \bar \nu_l$ and $Z\rightarrow l^+ l^-$.    $\nu\nu t\bar th$ has a negligible contribution in this channel, since the probability of  $h\rightarrow l^+ l^-$ is close to $0$.

The main background processes are 
\begin{eqnarray}
    &&\mu\mu\rightarrow \nu\bar \nu t_L\bar{t}_L/\nu\bar \nu t_R\bar{t}_RZ_L(\text{SM}) \rightarrow 2\nu2\bar\nu 2(l^+l^-) 2b\n\\
    &&\mu\mu\rightarrow t_L\bar{t}_L/t_R\bar t_R Z_L \rightarrow \nu\bar\nu 2(l^+l^-) 2b\n\\
    &&\mu\mu\rightarrow \nu\bar \nu Z_LZ_LZ_L \rightarrow  \nu\bar \nu 2(l^+l^-) 2b \n\\
    &&\mu\mu\rightarrow \nu\bar \nu Z_LZ_Lh \rightarrow  \nu\bar \nu 2(l^+l^-) 2b\\
    &&\mu\mu\rightarrow l^+\nu_lW^-Z_LZ_L(/l^-\bar\nu_l W^+Z_LZ_L) \rightarrow \nu\bar\nu 2(l^+l^-) 2b\n\\
    &&\mu\mu\rightarrow l^+\nu_lW^-Z_Lh(/l^-\bar\nu_l W^+Z_Lh) \rightarrow \nu\bar\nu 2(l^+l^-) 2b\n
\end{eqnarray}
of which the largest contribution to the background is $\nu\nu t\bar tZ$ from SM. Processes with negligible cross sections are not listed. We only gives the results of $30$ TeV, since the limits on $\delta y_t$ at $10$ TeV is too weak ($|C_{t\phi}|< 2 \ \text{TeV}^{-1}$).

\subsubsection*{30 TeV}
For all the processes, we implement a set of preliminary cuts:
\begin{equation}
  p_T(l) > 40 \ \text{GeV}, \ \   p_T(b) > 40 \ \text{GeV}
\end{equation}
with $b=b/\bar b$ and $l=l^+/l^-$.


The cuts we apply  on the events  are summarized as following:
\begin{itemize}\label{item:cuts_vvttz_lep_30}
    \item  Cut 1: Reject $80\  \text{GeV} <  M_{b_1 b_2} <  100 $ GeV; Reject $120\  \text{GeV} <  M_{b_1 b_2} <  130 $ GeV. The purpose of these cuts is to reduce the processes such as $\nu\bar \nu ZZZ$,  $\nu\bar \nu Z Z h$,  $l^+v_lW^+Z h$ that have $Z$ and $h$ decaying to $2b$.
    \item     Cut 2: Reject  $p_T(l) > 40 \ \text{GeV}$; Reject \ $p_T(b) > 40 \ \text{GeV}$. The purpose of these cuts is to reduce the SM part of signal process $\nu\bar \nu t\bar{t}Z$.
    \item Cut 3: Reject $\eta(l^+) < -3$ and $\eta(l^+) > 3$; Reject $\eta(b) < -3$ and $\eta(b) > 3$.  The purpose of these cuts is also to reduce the SM part of signal process.
\end{itemize}

\begin{figure}
    \centering
    \includegraphics[width=80mm]{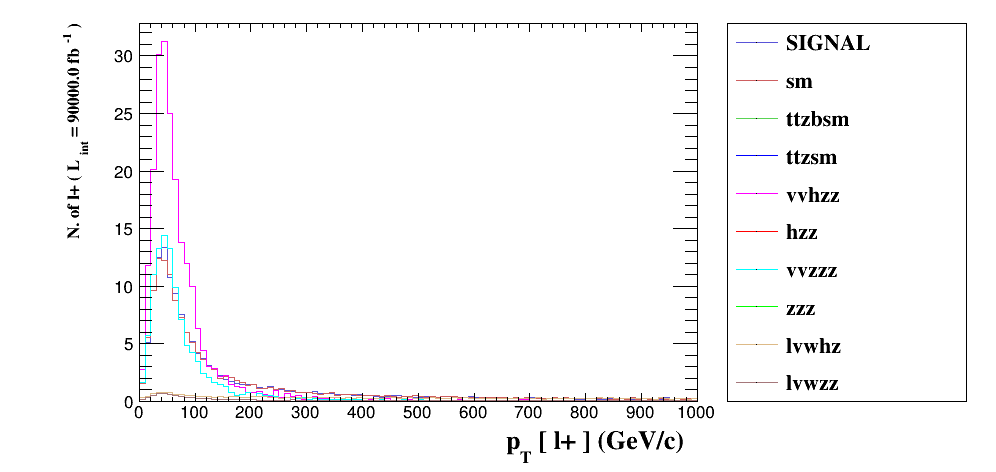}
    \includegraphics[width=80mm]{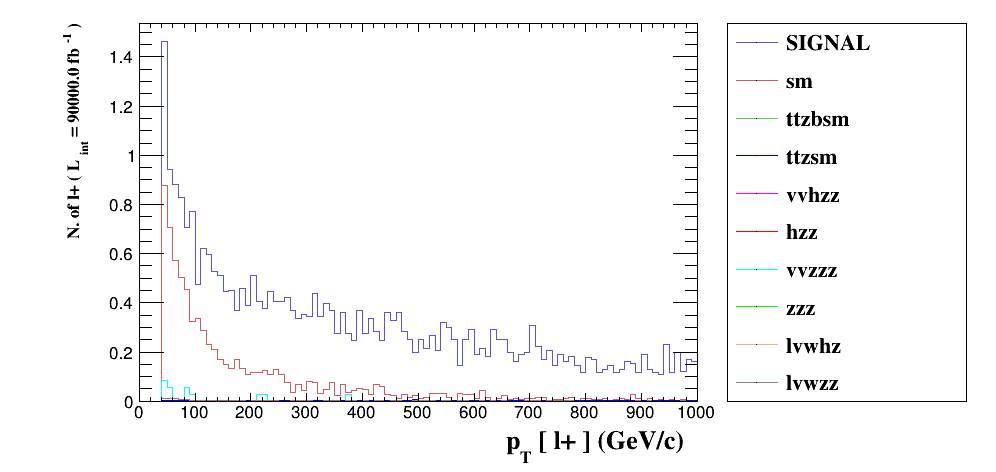}
    \includegraphics[width=80mm]{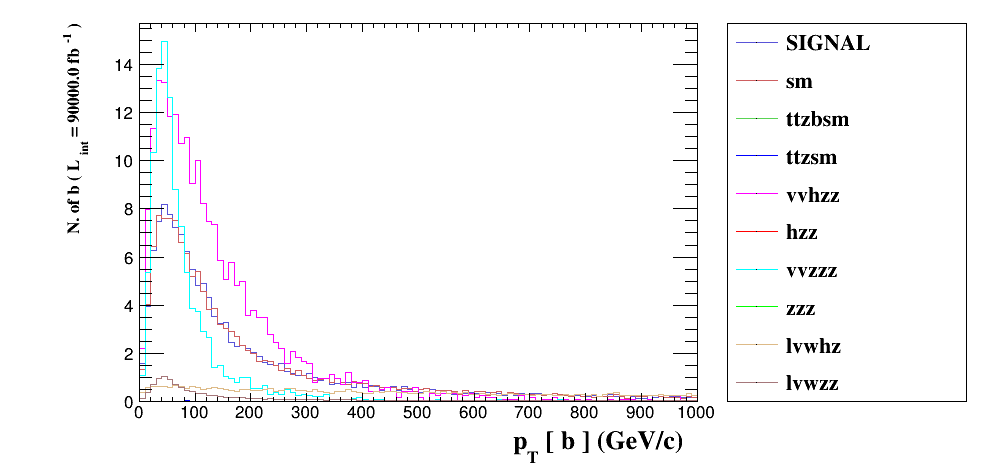}
    \includegraphics[width=80mm]{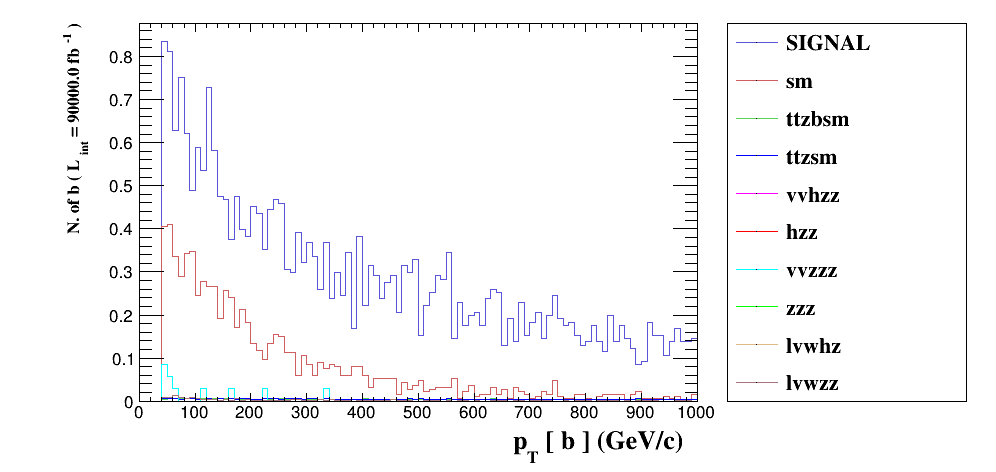}
    \includegraphics[width=80mm]{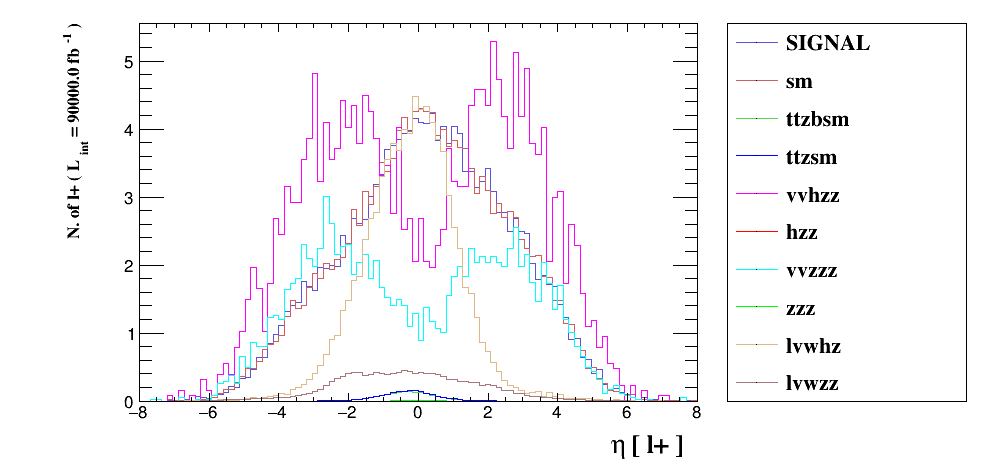}
    \includegraphics[width=80mm]{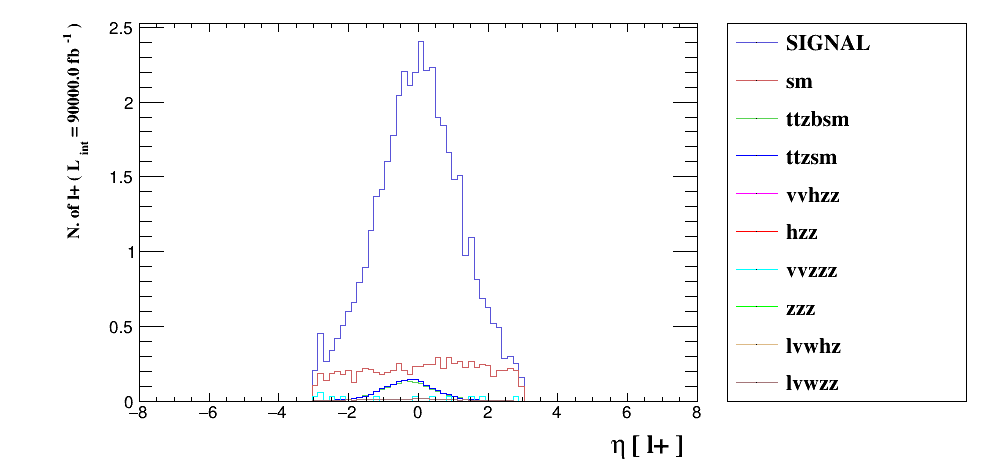}
    \includegraphics[width=80mm]{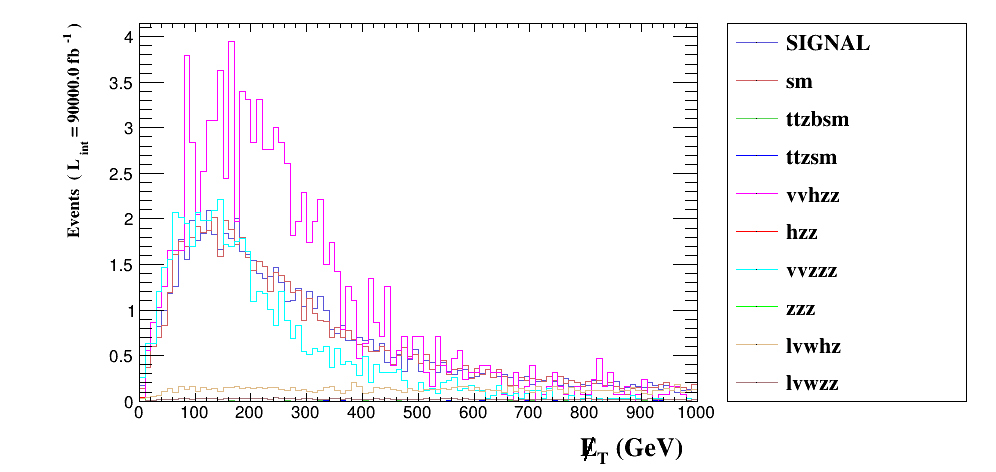}
    \includegraphics[width=80mm]{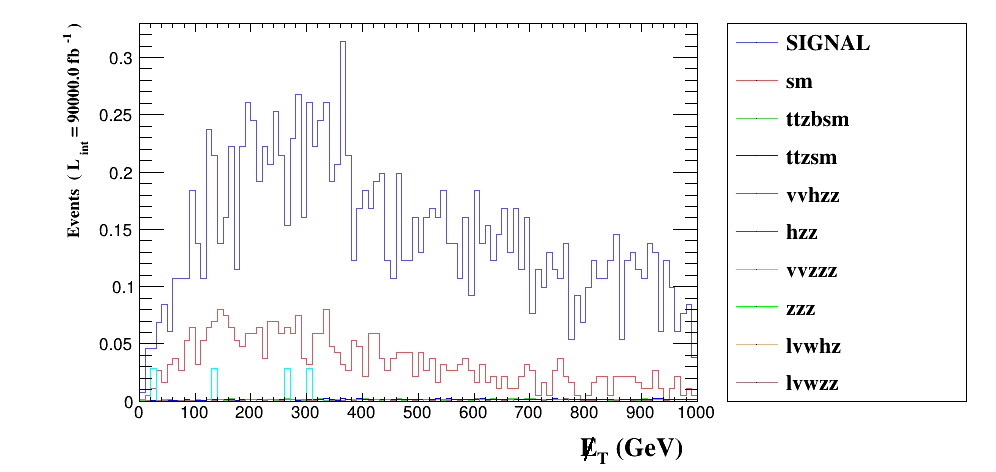}
    \includegraphics[width=80mm]{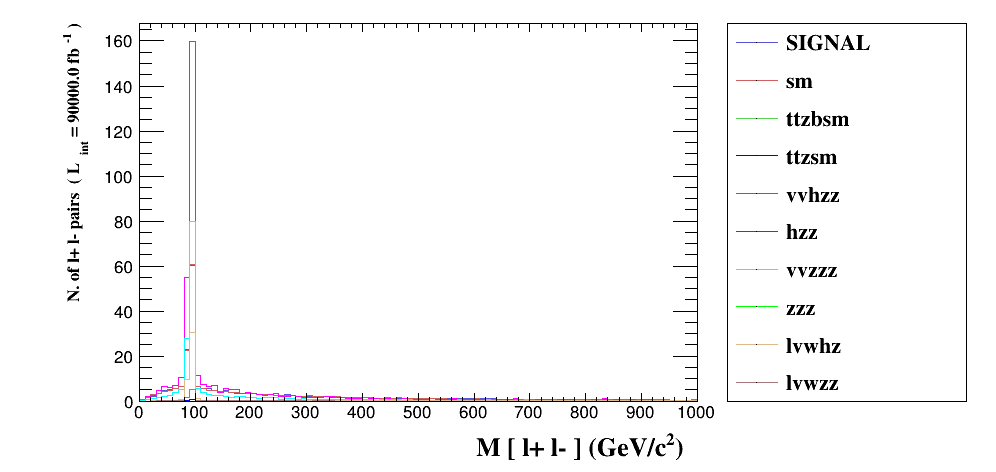}
    \includegraphics[width=80mm]{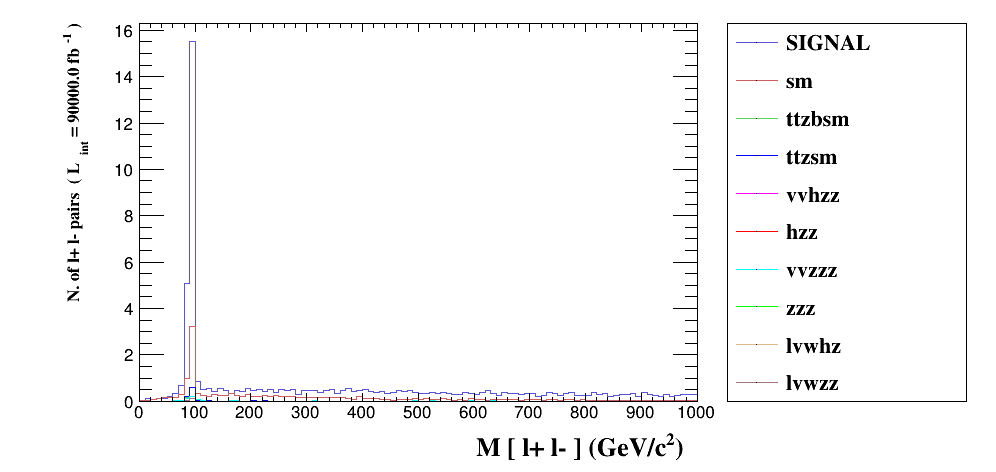} 
    \caption{Distributions of the lepton channel of $\nu\nu t \bar{t} Z$ at 30 TeV before(left) and after(right) cuts, see also Item(\ref{item:cuts_vvttz_lep_30}) for the cut and Table(\ref{tab:cutflow_vvttz_lep_30}) for the cut flow. }
    \label{fig:dist_vvttz_lep_30}
\end{figure}

The results of the cut flow are summarized in Table(\ref{tab:cutflow_vvttz_lep_30}), according to which, combined with b tagging efficiency $\epsilon_b=0.9$ and spin tagging efficiency $\epsilon_s$, we can obtain the corresponding statistical significance $\mathcal S$. Taking $\epsilon_s=0.9$ and following Eq.(\ref{eq:signficance}), we obtain $\mathcal S=6.0(\sigma)$ at $c_{t\phi}=2$ with $\Lambda = 1$ TeV. We also show some plots of the events before and after the cuts in Fig.(\ref{fig:dist_vvttz_lep_30}).  We also obtain the constraint on $c_{t\phi}$ with $\Lambda = 1$ TeV at $\epsilon_s=0.9$ as  $ -1.2\leq  c_{t\phi}\leq 1.1$ at $2\sigma$ and    $ -0.95\leq  c_{t\phi}\leq 0.9$ at $1\sigma$.  Converting them into constraints on $\delta y_t$, we get $ -6.7\% \leq  \delta y_t\leq 7.3\% $ at $2\sigma$ and    $ -5.5\%\leq  
\delta y_t\leq 5.8\%$ at $1\sigma$.

\begin{table}[]
    \centering
    \begin{tabular}{|c|c|c|c|c|}
    \hline
    \hline
     process                & before cuts & cut1 & cut2    & cut3 \\
     \hline
     BSM($\nu\bar{\nu} t_R\bar t_RZ_L/\nu\bar \nu t_L\bar t_L Z_L$)&  76.567$\times$2& 71.54$\times$2& 25.08$\times$2& 21.22$\times$2\\
    \hline 
    SM($\nu\bar{\nu} t_R\bar t_RZ_L/\nu\bar \nu t_L\bar t_L Z_L$)&  53.402$\times$2&  48.74$\times$2 &  7.26$\times$2& 3.97$\times$2\\
    \hline
    $t_R\bar t_RZ_L/ t_L\bar t_L Z_L$                             & 0.8574$\times$2&  0.8532$\times$2& $0.794\times 2$ &  0.785$\times$2\\
    \hline
    $\nu \bar{\nu} hZ_LZ_L$&    105.676&  0&  0&          0\\\hline
    \hline
    $\nu \bar{\nu} Z_LZ_LZ_L$&    53.399&    2.61&     0.402& 0.172\\\hline
    \hline
    $l\nu _lWZ_LZ_L$&    6.7635$\times$2&     0.362$\times$2&   0.177$\times$2& 0.151$\times$2\\
    \hline
    $l\nu _lWhZ_L$&    41.633$\times$2&    0&  0& 0\\
    \hline
    \end{tabular}
    \caption{Cut flow of the lepton channel of $\nu\nu t\bar tZ$ at 30 TeV, integrated luminosity is $90 \ ab^{-1}$. The BSM process is simulated at $c_{t\phi}=2$ (with $\Lambda = 1$ TeV). Helicity selections are shown in the table.  Background processes that have negligible contribution are not listed. The b tagging efficiency and spin tagging efficiency are set to be $\epsilon_b=\epsilon_s=1$. }
    \label{tab:cutflow_vvttz_lep_30}
\end{table}

\begin{figure}
    \centering
    \includegraphics[width=160mm]{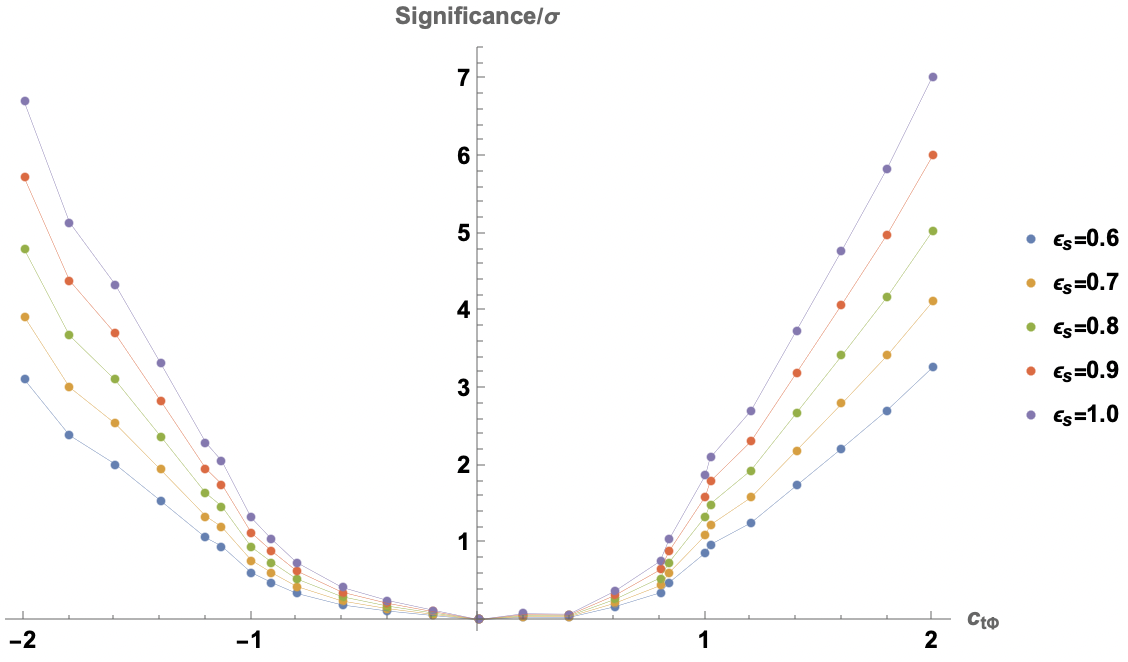}
    \caption{The dependence of statistical significance of the lepton channel of $\nu\nu t\bar tZ$ at 30 TeV on  $c_{t\phi}$ (with $\Lambda = 1$ TeV) and spin tagging efficiency $\epsilon_s$, the latter of $Z$ and $t/\bar t$  are take to be  equal. The b quark  tagging efficiency is set to be $\epsilon_b=0.9$.}
    \label{fig:sig_vvttz_lep_30}
\end{figure}

We analyze how statistical significance $\mathcal S$ changes with spin tagging efficiency $\epsilon_s$, as well as with  $c_{t\phi}$. For the values of $\epsilon_s$, we set them equal for  $Z$ and $t/\bar t$, then scan over $1, 0.9, 0.8, 0.7, 0.6$. The results are summarized in Fig.(\ref{fig:sig_vvttz_lep_30}).

\newpage
\subsection {Semi-lepton Channel}


In this subsection we study  channel $3l^{\pm}jj2b+\text{MET}$ with $l=e,\mu$, $b=b/\bar b$, which  comes  from $\nu\nu t\bar tZ$ through decays: $t\rightarrow b W^+\overset{W^+ \rightarrow l^+\nu_l}{\longrightarrow} b l^+ \nu_l$, $\bar t\rightarrow \bar b W^-\overset{W^- \rightarrow j j}{\longrightarrow} \bar b j j$ and  $Z\rightarrow l^+ l^-$, or with $t(l^+)\leftrightarrow \bar t(l^-)$. 

The main background processes are
\begin{eqnarray}
    &&\mu\mu\rightarrow \nu\bar \nu t_R\bar{t}_R(/t_L\bar{t}_L)Z_L(\text{SM}) \rightarrow  3\nu l^{\pm} 2bjj\n\\
    &&\mu\mu\rightarrow t_R\bar t_R(/t_L\bar{t}_L) Z_L \rightarrow  3\nu l^{\pm} 2bjj\n\\
    &&\mu\mu\rightarrow l^+\nu_lW^-Z_LZ_L(/l^-\bar\nu_l W^+Z_LZ_L) \rightarrow \nu 3l^{\pm}2bjj\\
    &&\mu\mu\rightarrow l^+\nu_lW^-Z_Lh(/l^-\bar\nu_l W^+Z_Lh) \rightarrow \nu 3l^{\pm}2bjj\n\\
    &&\mu\mu\rightarrow l^+\nu_lW^-hh(/l^-\bar\nu_l W^+hh) \rightarrow \nu 3l^{\pm}2bjj\n
\end{eqnarray}

We will only give the results of $30$ TeV, since the limits on $\delta y_t$ at $10$ TeV is too weak ($|C_{t\phi}|< 2 \ \text{TeV}^{-1}$).


\subsubsection*{30 \text{TeV}}

The semi-lepton channel of $\nu\bar{\nu} t_R\bar{t}_R(/t_L\bar t_L) Z_L$  has a much larger cross-section than the lepton channel, even though the ratio of the signal cross sections between the semi-lepton and lepton is not as prominent as in $\nu\bar{\nu} t_R\bar{t}_R(/t_L\bar t_L) h$, we still expect a higher significance. Moreover, we will perform some cuts to reduce the background and enhance the statistical significance of the processes $\nu\bar{\nu} t_R\bar{t}_R Z_L$ and $\nu\bar{\nu} t_L\bar{t}_L Z_L$. The cuts are listed as below:

\begin{itemize}\label{item:cuts_vvttz_semi_30}
    \item  Cut 1: Select  $p_T(l) > 40 \ \text{GeV}$. The purpose of this cut is to reduce the SM part of signal process $\nu\bar{\nu} t_R\bar{t}_R(/t_L\bar t_L) Z_L$
    \item  Cut 2: Reject $80\  \text{GeV} < M_{b_1 b_2} < 100\  \text{GeV}$. The purpose of this cut is to reduce the process $l^+\nu_lW^-Z_LZ_L(/l^-\bar\nu_l W^+Z_LZ_L)$ that have $Z$ decaying to $2b$.
    \item Cut 3: Reject $120\  \text{GeV} < M_{b_1 b_2} < 130\  \text{GeV}$. The purpose of this cut is to reduce the process $l^+\nu_lW^-Z_Lh(/l^-\bar\nu_l W^+Z_Lh)$ that have $h$ decaying to $2b$.
\end{itemize}
with $b=b/\bar b$, $l=l^+l^-$.

\begin{table}[]
    \centering
    \begin{tabular}{|c|c|c|l|l|}
    \hline
    \hline
     process                & before cuts & cut1& cut2&cut3\\
     \hline
     BSM($\nu\bar{\nu} ttZ_L$)&  229.7$\times$4& 228.16$\times$4&218.19$\times$4&213.09$\times$4\\
    \hline 
    SM($\nu\bar{\nu} ttZ_L$)&  160.2$\times$4&  158.65$\times$4&150.31$\times$4&145.61$\times$4\\\hline      
    \hline
    $ttZ_L$&    2.57$\times$4&  2.57$\times$4&2.56$\times$4&2.56$\times$4\\\hline
    \hline
    $l\nu _lWZ_LZ_L$&    20.23$\times$2&   20.23$\times$2&1.16$\times$2&1.12$\times$2\\
    \hline
    $l\nu _lWhZ_L$&    124.06$\times$2&  124.06$\times$2&124.06$\times$2&0\\\hline
    \end{tabular}
    \caption{Cut flow of semi-lepton channel of $\nu\nu t\bar tZ$ at 30 TeV, integrated luminosity is $90 \ ab^{-1}$.  The BSM process is simulated at $c_{t\phi}=2$ (with $\Lambda = 1$ TeV). Helicity selections are shown in the table.  Background processes that have negligible contribution are not listed. The b tagging efficiency and spin tagging efficiency are set to be $\epsilon_b=\epsilon_s=1$. }
    \label{tab:cutflow_vvttz_semi_30}
\end{table}

\begin{figure}
    \centering
    \includegraphics[width=80mm]{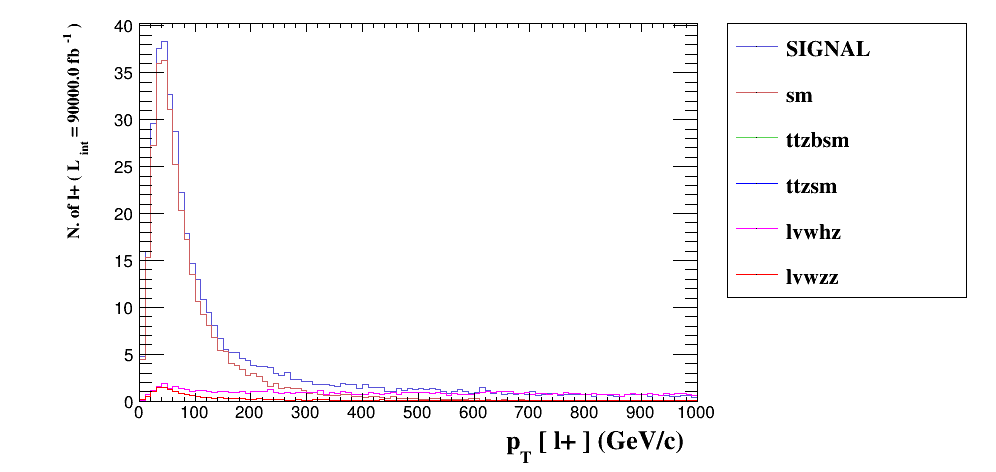}
    \includegraphics[width=80mm]{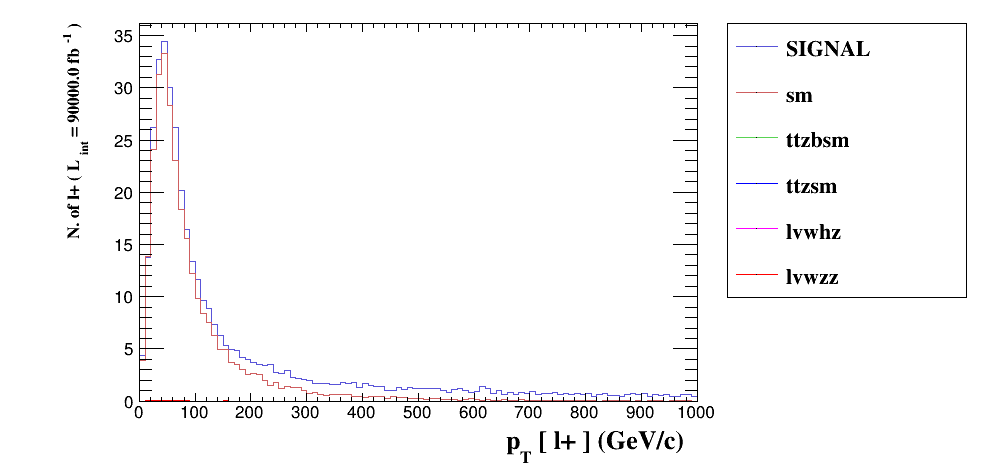}
    \includegraphics[width=80mm]{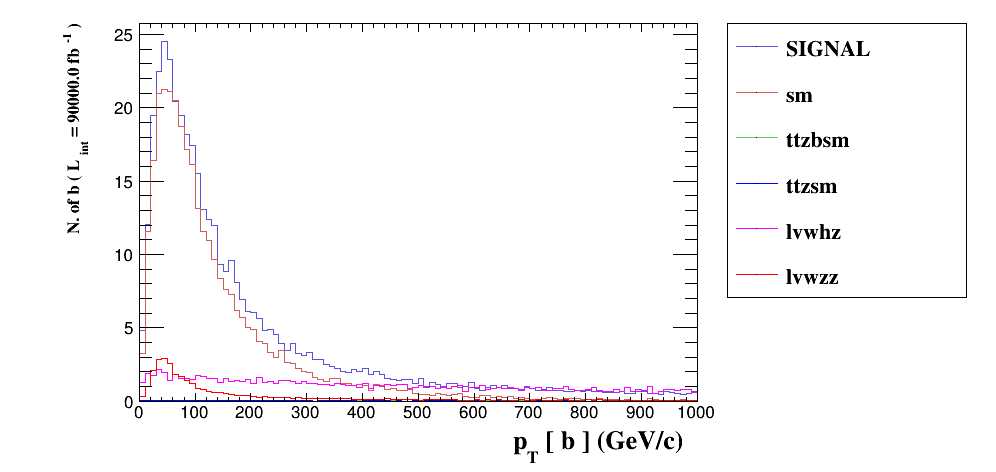}
    \includegraphics[width=80mm]{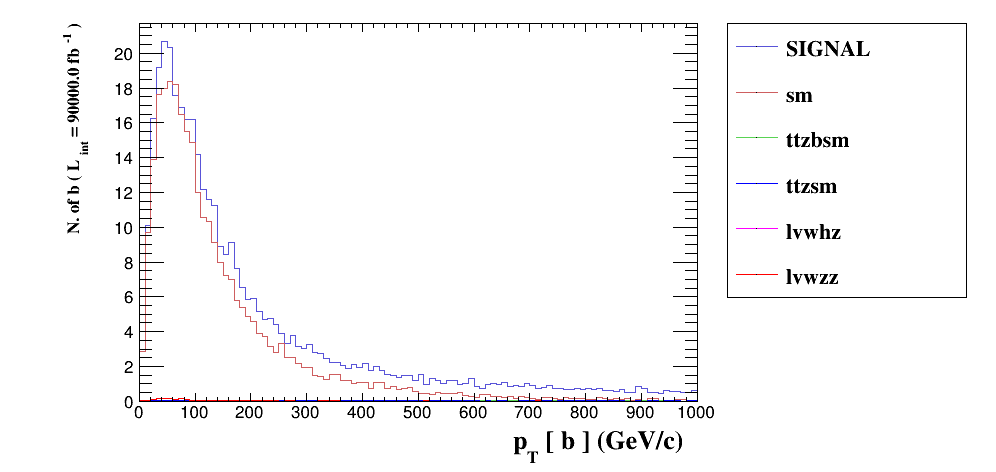}
    \includegraphics[width=80mm]{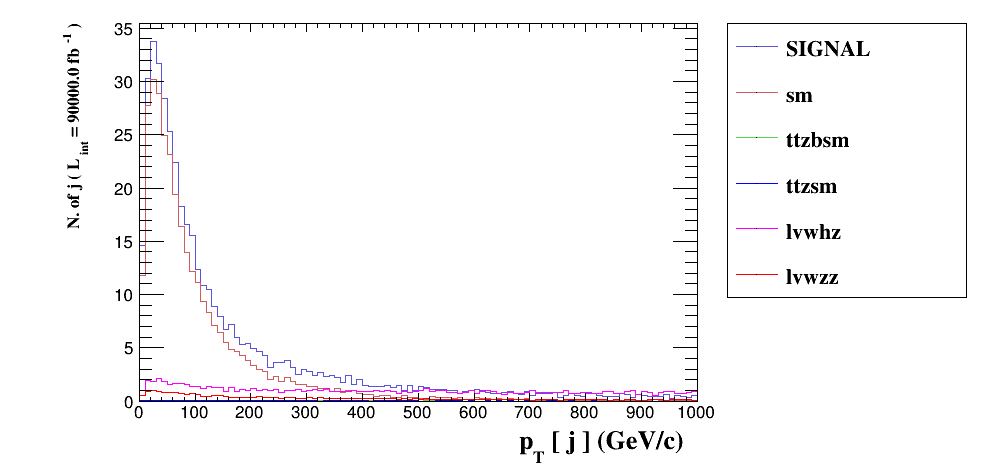}
    \includegraphics[width=80mm]{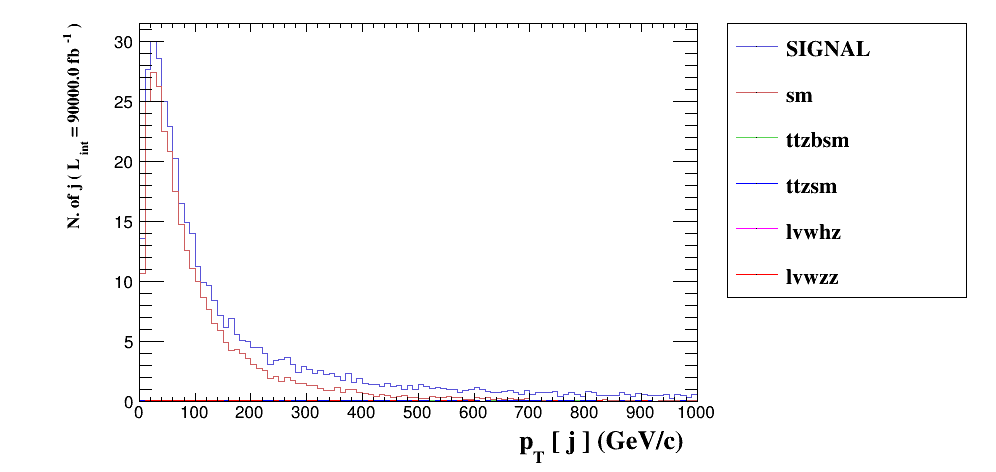}
    \includegraphics[width=80mm]{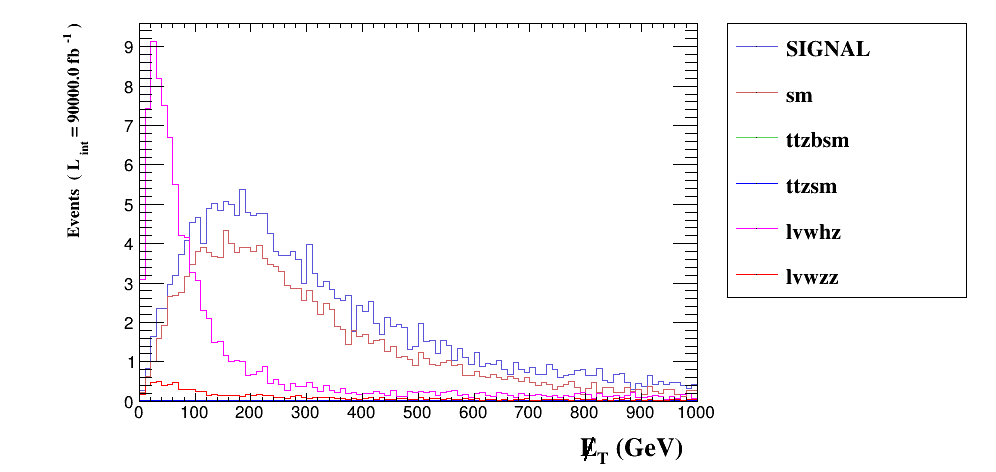}
    \includegraphics[width=80mm]{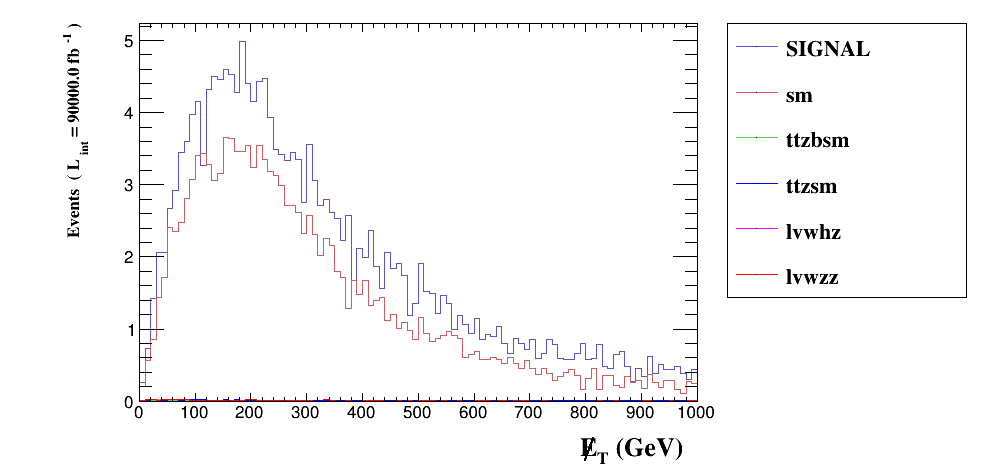}
    \includegraphics[width=80mm]{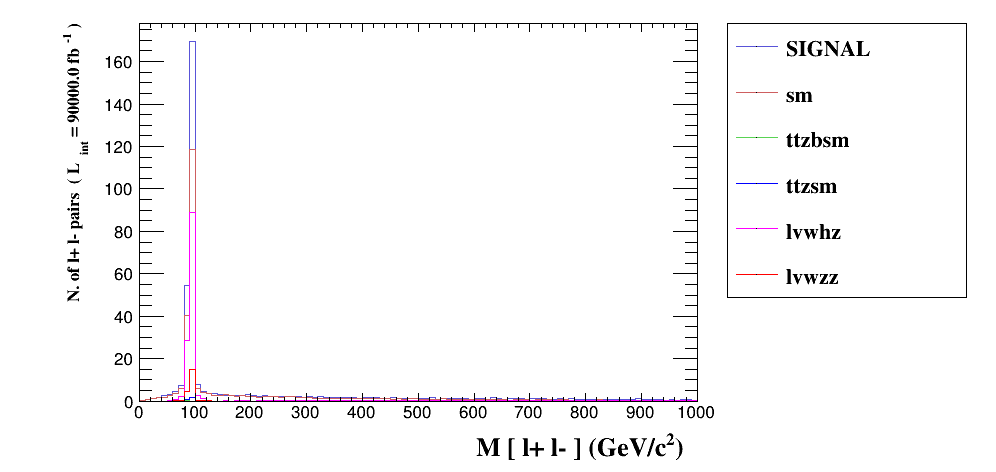}
    \includegraphics[width=80mm]{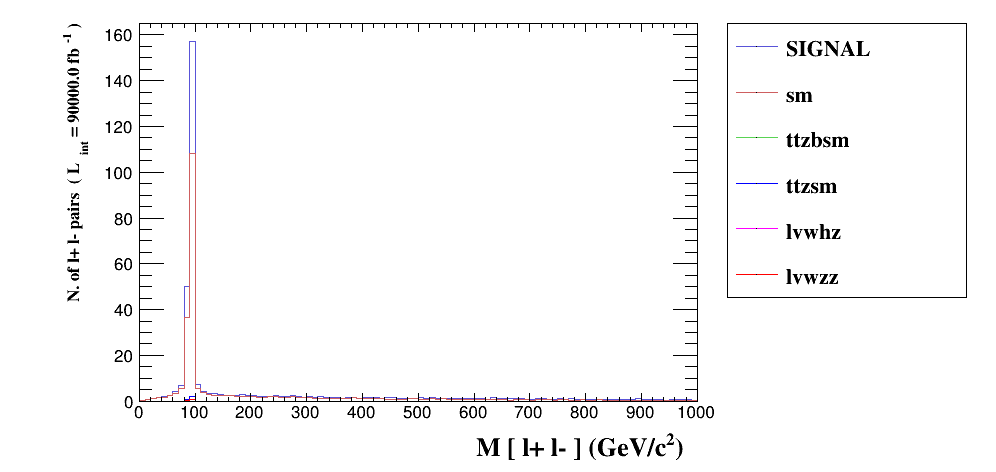}  
\caption{Distributions of the semi-lepton channel of $\nu\bar{\nu}t\bar{t}Z$ at 30 TeV before(left) and after(right) cuts, see Item(\ref{item:cuts_vvttz_semi_30}) for the cuts and  Table(\ref{tab:cutflow_vvttz_semi_30}) for event numbers with the cut flow. }
    \label{fig:dist_vvttz_semi_30}
\end{figure}

The results of the cut flow are summarized in Table(\ref{tab:cutflow_vvttz_semi_30}), according to which, combined with b tagging efficiency $\epsilon_b=0.9$ and spin tagging efficiency $\epsilon_s$, we can obtain the corresponding statistical significance. Taking $\epsilon_s=0.9$ and following Eq.(\ref{eq:signficance}), we obtain $\mathcal S=7.6 (\sigma)$ at $c_{t\phi} =2$ with $\Lambda = 1$ TeV. We also show some plots of the events before and after the cuts in Fig.(\ref{fig:dist_vvttz_semi_30}). Following the same procedure we can obtain the constraint on $C_{t\phi}$ with $\epsilon_s=0.9$ as   $-1.18\  \text{TeV}^{-2}\leq  C_{t\phi}\leq 1.17 \ \text{TeV}^{-2}$ at $2\sigma$ and    $-0.98\  \text{TeV}^{-2}\leq  C_{t\phi}\leq 0.92\  \text{TeV}^{-2}$ at $1\sigma$.  Converting them into constraints on $\delta y_t$, we get $-7.2\% \leq  \delta y_t\leq 7.2\%$ at $2\sigma$ and    $-5.6\%\leq  \delta y_t\leq 6.0\%$ at $1\sigma$.

\begin{figure}
    \centering
    \includegraphics[width=160mm]{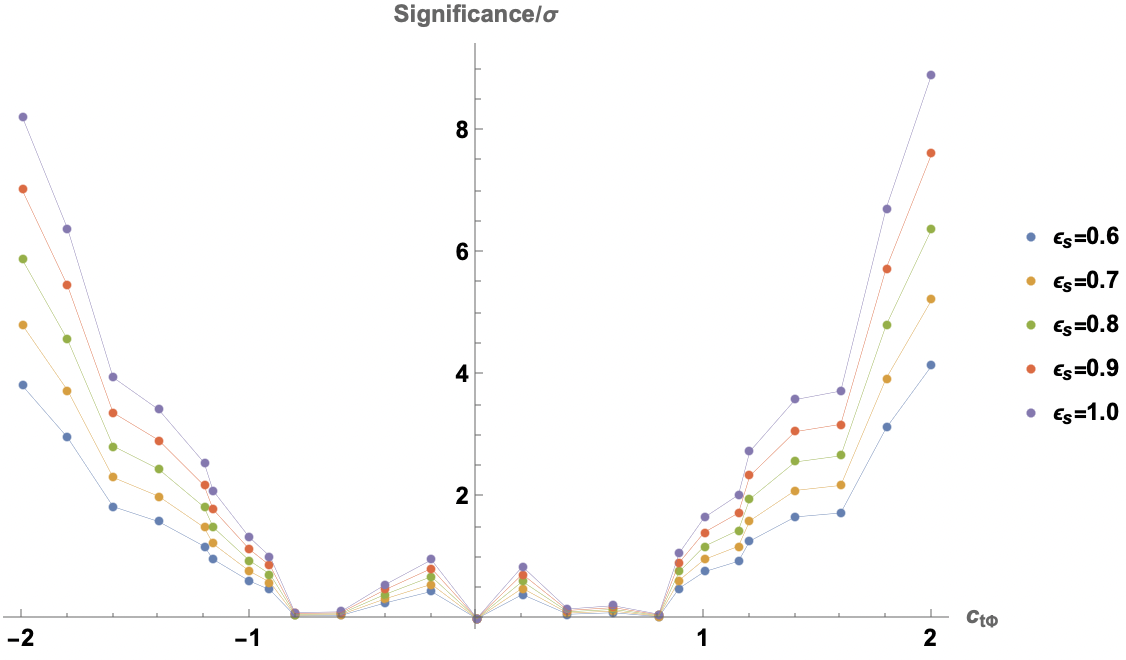}
    \caption{The dependence of statistical significance of the semi-lepton channel of $\nu\nu t\bar tZ$ at 30 TeV on  $c_{t\phi}$ (with $\Lambda = 1$ TeV) and spin tagging efficiency $\epsilon_s$, the latter of $Z$ and $t/\bar t$  are take to be  equal. The b quark  tagging efficiency is set to be $\epsilon_b=0.9$.}
    \label{fig:sig_vvttz_semi_30}
\end{figure}

Statistical significance $\mathcal S$ also changes with spin tagging efficiency $\epsilon_s$, as well as with  $c_{t\phi}$. The results are summarized in Fig.(\ref{fig:sig_vvttz_semi_30}).

\pagebreak
\section{Conclusion}
\label{sec:con}

In this paper, we studied the measurement of the top Yukawa coupling through $2\rightarrow 3 $ VBS  in future muon colliders. We focused on the final states of $\nu\nu t\bar th$ and $\nu\nu t\bar tZ$, corresponding to  $W^+W^-\rightarrow t\bar t h/Z$. 

By analysing the related  helicity amplitudes and hard processes without decaying, we established the high sensitivity of the processes to the anomalous top Yukawa coupling $\delta y_t$. This sensitivity is particularly enhanced by selecting the helicities of $t\bar t$ and $Z$ to be $t_R\bar t_R/t_L\bar t_L$ and $Z_L$.  
We also obtain the limits  on $\delta y_t$ for hard processes(Table(\ref{tab:constrt_nodecay})) as  rough estimations.  As can be seen in the table, the limits from $\nu\nu t\bar th$ and $\nu\nu t\bar tZ$ are close, with the former better at the positive limits and the latter better at the negative limits. The results from $\nu\nu t\bar th+\nu\nu t\bar tZ$ combined are better than both of the two individually. At $30$ TeV, it gives the limit on $\delta y_t$ of $[-0.36\%, 0.92\%]$ at $1\sigma$ level. 

 We then proceed to carry out detailed background analysis of the processes after decaying  at  $10$ TeV ($10  \  \text{ab}^{-1}$) and $30$ TeV($90 \ \text{ab}^{-1}$) respectively, with the main focus on the lepton and semi-lepton channels.  The results are summarized in Table(\ref{tab:constraints_summary}). For an overall picture,  the results from $30$ TeV are significantly better than  the counter parts from $10$ TeV; $\nu\nu t\bar t h$ is generally better than $\nu\nu t\bar tZ$, but the two are still at the same level;  the semi-lepton channels are consistently better than lepton channels, but the differences are smaller than between $\nu\nu t\bar t h$ and $\nu\nu t\bar t Z$; finally, the variation of the results from changing  spin tagging efficiency is minimal.  As the most stringent limit on $\delta y_t$, at $30$ TeV and $\epsilon_s=0.9$,  the semi-lepton channels of $\nu\nu t\bar th$ and $\nu\nu t\bar tZ$  constrain $\delta y_t$ at  $[-1.8\%, 1.6\%]$ and $[-5.6\%, 6.0\%]$ at $1\sigma$.
 
 \begin{table}[]
    \centering
    \begin{tabular}{|c|c|c|c|c|c|}
    \hline
    \hline
\multicolumn{2}{|c|}{Channels}      & $E_{cm}$ & spin tagging          & \multicolumn{2}{|c|}{Limits of $\delta y_t$} \\
\cline{5-6}
       \multicolumn{2}{|c|}{ }         &           &efficiency($\epsilon_s$)  &  $1\sigma$  & $2\sigma$      \\ 
    \hline
    \multirow{6}{*}{$\nu\nu t\bar th$}
   & \multirow{3}{*}{lepton}  & \multirow{2}{*}{30 TeV}     &  0.9      &  $[-3.1\%, 2.6\%]$      &      $[-3.2\%,3.4\%]$  \\
            &                  &                           &  0.7     &  $[-3.2\%, 3.1\%]$       &      $[-3.7\% , 3.8\%]$     \\
    \cline{3-6}
            &                &   \multirow{2}{*}{10 TeV}  &   0.9   &    $[-11.0\% , 9.8\%]$     & -- \\
            &                &                           &  0.7     &  $[-12.2\%, 11.0\%]$       &   -- \\     
    \cline{2-6}      
    & \multirow{3}{*}{semi-lepton} &   \multirow{2}{*}{30 TeV}  &   0.9     & $[-1.6\%, 1.8\%]$     &    $[-2.4\% , 2.7\%]$    \\
    &                             &                             &   0.7     & $[-2.1\%, 2.4]$      &     $[-2.5\% , 2.8\%]$ %
     \\  
    \cline{3-6}
    &                            &   \multirow{2}{*}{10 TeV}    &   0.9     &  $[-7.0\%, 6.7\%]$   &     $[-9.8\%, 9.8\%]$  \\
    &                            &                              &   0.7     &  $[-8.3\%, 7.9\%]$   &     $[-11.2\%, 10.9\%]$  \\

    \hline 
   \multirow{6}{*}{$\nu\nu t\bar tZ$}
   & \multirow{3}{*}{lepton}      &   \multirow{2}{*}{30 TeV}   &  0.9      &   $[-5.5\%, 5.8\%]$    &   $[-6.7\% , 7.3\%]$       \\
   &                              &                            &  0.7        &  $[-5.8\%, 6.6\%]$    &    $[-8.3\%, 8.6\%]$  \\
     \cline{3-6}        
   &           &    10 TeV  &   --         &     --       & -- \\    
     \cline{2-6} 
   &\multirow{3}{*}{semi-lepton} &    \multirow{2}{*}{30 TeV}    &  0.9     &   $[-5.6\%, 6.0\%]$      &      $[-7.2\%, 7.2\%]$   \\ 
   &                            &                                &  0.7     &  $[-6.1\%, 7.2\%]$    &    $[-8.3\%, 8.6\%]$  \\
    \cline{3-6}
   &           &    10 TeV   &   0.9        &        --   &  --  \\ 
    \hline                
    \end{tabular}
    \caption{Summary of constraints on $\delta y_t$ in different channels with different settings. b tagging efficiency is taken to be $0.9$ for every analysis. The helicities of $t\bar t $ are chosen to be $LL/RR$, the helicity of $Z$ is chosen to be longitudinal. Spin tagging efficiency of final states are chosen be the same.   }
    \label{tab:constraints_summary}
\end{table}


Our results demonstrate promising prospects and importance for $2\rightarrow 3$ VBS to measure the top Yukawa coupling for futhure muon colliders at 10 TeV and 30 TeV, especially the latter. In particular, the important processes include not only $\nu\nu t \bar t h$, but also $\nu\nu t\bar t Z$. The latter process has never been studied before this paper as far as we're concerned. 
 Our analysis (of hard processes) nevertheless show comparable results with those using bin-by-bin analysis. 
The crucial setting of our approach is to specify the helicities of final states, which dramatically enhanced significance. The spin of a heavy particle has been measured in LHC several times with high precision\cite{Ballestrero:2020qgv, De:2020iwq}, thus it's reasonable to assume that spin measurement can serve a reliable tool for analysis in future muon colliders. However, further study and improvement is obviously needed, and indeed  has potential of bearing great fruit.  Finally, we did a comprehensive background analysis with all lepton channels and semi-lepton channels of $\nu\nu t\bar th$ and $\nu\nu t\bar t Z$. The best limits on $\delta y_t$ are  worse than hard processes, but the deviations are not very large.  It seems that the main factor of the decrease of statistical significance is the smaller event numbers after decaying the heavy particles, not the effects of background processes.  

%

This paper is also limited in several ways. First of all, we only studied $\nu\nu t\bar t h/Z$ out of all $2\rightarrow 3$ VBS processes(Eq.(\ref{eq:processes})),  also didn't include the hadronic channels of $\nu\nu t\bar t h/Z$ which have the largest cross sections. 
Second, our main method in data analysis is still to apply cuts to reduce background, apart from helicity selection. Bin-by-bin  analysis, which makes use of the differential cross sections more efficiently, could greatly enhance the constraints on $y_t$.   However, the two methods are not mutually exclusive, it should be possible to combine spin tagging setting with bin-by-bin analysis. This would make the best use of the data and put strongest constraints on top Yukawa. To finally conclude, measuring  top Yukawa coupling with $2\rightarrow 3$ VBS processes has shown great prospects and this paper is only a beginning, it's worth further  study and  investigation.

\section*{Acknowledgements}

We would like to thank Yang Ma for helpful suggestions. This paper  is supported by  NSFC (National Natural Science Foundation of China) no. 12205118.

\bibliographystyle{JHEP}
\bibliography{references-top}

\providecommand{\href}[2]{#2}\begingroup\raggedright\begin{thebibliography}{10}

\bibitem{KAPLAN1984187}
D.B.~Kaplan, H.~Georgi and S.~Dimopoulos, \emph{Composite higgs scalars},
  \href{https://doi.org/https://doi.org/10.1016/0370-2693(84)91178-X}{\emph{Physics
  Letters B} {\bfseries 136} (1984) 187}.

\bibitem{DUGAN1985299}
M.J.~Dugan, H.~Georgi and D.B.~Kaplan, \emph{Anatomy of a composite higgs
  model},
  \href{https://doi.org/https://doi.org/10.1016/0550-3213(85)90221-4}{\emph{Nuclear
  Physics B} {\bfseries 254} (1985) 299}.

\bibitem{Bally:2022naz}
A.~Bally, Y.~Chung and F.~Goertz, \emph{{Hierarchy problem and the top Yukawa
  coupling: An alternative to top partner solutions}},
  \href{https://doi.org/10.1103/PhysRevD.108.055008}{\emph{Phys. Rev. D}
  {\bfseries 108} (2023) 055008}
  [\href{https://arxiv.org/abs/2211.17254}{{\ttfamily 2211.17254}}].

\bibitem{Trodden:1998ym}
M.~Trodden, \emph{{Electroweak baryogenesis}},
  \href{https://doi.org/10.1103/RevModPhys.71.1463}{\emph{Rev. Mod. Phys.}
  {\bfseries 71} (1999) 1463}
  [\href{https://arxiv.org/abs/hep-ph/9803479}{{\ttfamily hep-ph/9803479}}].

\bibitem{Zhang:1994fb}
X.~Zhang, S.K.~Lee, K.~Whisnant and B.L.~Young, \emph{{Phenomenology of a
  nonstandard top quark Yukawa coupling}},
  \href{https://doi.org/10.1103/PhysRevD.50.7042}{\emph{Phys. Rev. D}
  {\bfseries 50} (1994) 7042}
  [\href{https://arxiv.org/abs/hep-ph/9407259}{{\ttfamily hep-ph/9407259}}].

\bibitem{Baer:2023uwo}
H.~Baer, V.~Barger, J.~Dutta, D.~Sengupta and K.~Zhang, \emph{{Top squarks from
  the landscape at high luminosity LHC}},
  \href{https://doi.org/10.1103/PhysRevD.108.075027}{\emph{Phys. Rev. D}
  {\bfseries 108} (2023) 075027}
  [\href{https://arxiv.org/abs/2307.08067}{{\ttfamily 2307.08067}}].

\bibitem{ATLAS:2024zkx}
{\scshape ATLAS} collaboration, \emph{{Search for R-parity violating
  supersymmetric decays of the top squark to a b-jet and a lepton in
  s=13\,\,TeV pp collisions with the ATLAS detector}},
  \href{https://doi.org/10.1103/PhysRevD.110.092004}{\emph{Phys. Rev. D}
  {\bfseries 110} (2024) 092004}
  [\href{https://arxiv.org/abs/2406.18367}{{\ttfamily 2406.18367}}].

\bibitem{ATLAS:2024vyj}
{\scshape ATLAS} collaboration, \emph{{A search for top-squark pair production,
  in final states containing a top quark, a charm quark and missing transverse
  momentum, using the 139 fb$^{−1}$ of pp collision data collected by the
  ATLAS detector}}, \href{https://doi.org/10.1007/JHEP07(2024)250}{\emph{JHEP}
  {\bfseries 07} (2024) 250}
  [\href{https://arxiv.org/abs/2402.12137}{{\ttfamily 2402.12137}}].

\bibitem{ATLAS:2023cbt}
{\scshape ATLAS} collaboration, \emph{{Probing the CP nature of the
  top\textendash{}Higgs Yukawa coupling in tt\textasciimacron{}H and tH events
  with H\textrightarrow{}bb\textasciimacron{} decays using the ATLAS detector
  at the LHC}},
  \href{https://doi.org/10.1016/j.physletb.2024.138469}{\emph{Phys. Lett. B}
  {\bfseries 849} (2024) 138469}
  [\href{https://arxiv.org/abs/2303.05974}{{\ttfamily 2303.05974}}].

\bibitem{CMS:2020cga}
{\scshape CMS} collaboration, \emph{{Measurements of $\mathrm{t\bar{t}}H$
  Production and the CP Structure of the Yukawa Interaction between the Higgs
  Boson and Top Quark in the Diphoton Decay Channel}},
  \href{https://doi.org/10.1103/PhysRevLett.125.061801}{\emph{Phys. Rev. Lett.}
  {\bfseries 125} (2020) 061801}
  [\href{https://arxiv.org/abs/2003.10866}{{\ttfamily 2003.10866}}].

\bibitem{CMS:2020djy}
{\scshape CMS} collaboration, \emph{{Measurement of the top quark Yukawa
  coupling from $\mathrm{t\bar{t}}$ kinematic distributions in the dilepton
  final state in proton-proton collisions at $\sqrt{s}=$ 13 TeV}},
  \href{https://doi.org/10.1103/PhysRevD.102.092013}{\emph{Phys. Rev. D}
  {\bfseries 102} (2020) 092013}
  [\href{https://arxiv.org/abs/2009.07123}{{\ttfamily 2009.07123}}].

\bibitem{Delahaye:2019omf}
J.P.~Delahaye, M.~Diemoz, K.~Long, B.~Mansouli\'e, N.~Pastrone, L.~Rivkin
  et~al., \emph{{Muon Colliders}},
  \href{https://arxiv.org/abs/1901.06150}{{\ttfamily 1901.06150}}.

\bibitem{Long:2020wfp}
K.~Long, D.~Lucchesi, M.~Palmer, N.~Pastrone, D.~Schulte and V.~Shiltsev,
  \emph{{Muon colliders to expand frontiers of particle physics}},
  \href{https://doi.org/10.1038/s41567-020-01130-x}{\emph{Nature Phys.}
  {\bfseries 17} (2021) 289}
  [\href{https://arxiv.org/abs/2007.15684}{{\ttfamily 2007.15684}}].

\bibitem{Buttazzo:2018qqp}
D.~Buttazzo, D.~Redigolo, F.~Sala and A.~Tesi, \emph{{Fusing Vectors into
  Scalars at High Energy Lepton Colliders}},
  \href{https://doi.org/10.1007/JHEP11(2018)144}{\emph{JHEP} {\bfseries 11}
  (2018) 144} [\href{https://arxiv.org/abs/1807.04743}{{\ttfamily
  1807.04743}}].

\bibitem{Costantini:2020stv}
A.~Costantini, F.~De~Lillo, F.~Maltoni, L.~Mantani, O.~Mattelaer, R.~Ruiz
  et~al., \emph{{Vector boson fusion at multi-TeV muon colliders}},
  \href{https://doi.org/10.1007/JHEP09(2020)080}{\emph{JHEP} {\bfseries 09}
  (2020) 080} [\href{https://arxiv.org/abs/2005.10289}{{\ttfamily
  2005.10289}}].

\bibitem{Han:2020pif}
T.~Han, D.~Liu, I.~Low and X.~Wang, \emph{{Electroweak couplings of the Higgs
  boson at a multi-TeV muon collider}},
  \href{https://doi.org/10.1103/PhysRevD.103.013002}{\emph{Phys. Rev. D}
  {\bfseries 103} (2021) 013002}
  [\href{https://arxiv.org/abs/2008.12204}{{\ttfamily 2008.12204}}].

\bibitem{Chiesa:2020awd}
M.~Chiesa, F.~Maltoni, L.~Mantani, B.~Mele, F.~Piccinini and X.~Zhao,
  \emph{{Measuring the quartic Higgs self-coupling at a multi-TeV muon
  collider}}, \href{https://doi.org/10.1007/JHEP09(2020)098}{\emph{JHEP}
  {\bfseries 09} (2020) 098}
  [\href{https://arxiv.org/abs/2003.13628}{{\ttfamily 2003.13628}}].

\bibitem{Whisnant:1994fh}
K.~Whisnant, B.-L.~Young and X.~Zhang, \emph{{Unitarity and anomalous top quark
  Yukawa couplings}},
  \href{https://doi.org/10.1103/PhysRevD.52.3115}{\emph{Phys. Rev. D}
  {\bfseries 52} (1995) 3115}
  [\href{https://arxiv.org/abs/hep-ph/9410369}{{\ttfamily hep-ph/9410369}}].

\bibitem{Henning:2018kys}
B.~Henning, D.~Lombardo, M.~Riembau and F.~Riva, \emph{{Measuring Higgs
  Couplings without Higgs Bosons}},
  \href{https://doi.org/10.1103/PhysRevLett.123.181801}{\emph{Phys. Rev. Lett.}
  {\bfseries 123} (2019) 181801}
  [\href{https://arxiv.org/abs/1812.09299}{{\ttfamily 1812.09299}}].

\bibitem{Maltoni:2019aot}
F.~Maltoni, L.~Mantani and K.~Mimasu, \emph{{Top-quark electroweak interactions
  at high energy}}, \href{https://doi.org/10.1007/JHEP10(2019)004}{\emph{JHEP}
  {\bfseries 10} (2019) 004}
  [\href{https://arxiv.org/abs/1904.05637}{{\ttfamily 1904.05637}}].

\bibitem{Chen:2022yiu}
M.~Chen and D.~Liu, \emph{{Top Yukawa coupling measurement at the muon
  collider}}, \href{https://doi.org/10.1103/PhysRevD.109.075020}{\emph{Phys.
  Rev. D} {\bfseries 109} (2024) 075020}
  [\href{https://arxiv.org/abs/2212.11067}{{\ttfamily 2212.11067}}].

\bibitem{Liu:2023yrb}
Z.~Liu, K.-F.~Lyu, I.~Mahbub and L.-T.~Wang, \emph{{Top Yukawa coupling
  determination at high energy muon collider}},
  \href{https://doi.org/10.1103/PhysRevD.109.035021}{\emph{Phys. Rev. D}
  {\bfseries 109} (2024) 035021}
  [\href{https://arxiv.org/abs/2308.06323}{{\ttfamily 2308.06323}}].

\bibitem{Barman:2022pip}
R.K.~Barman et~al., \emph{{Directly Probing the CP-structure of the Higgs-Top
  Yukawa at HL-LHC and Future Colliders}},  in \emph{{Snowmass 2021}}, 3, 2022
  [\href{https://arxiv.org/abs/2203.08127}{{\ttfamily 2203.08127}}].

\bibitem{Cassidy:2023lwd}
M.E.~Cassidy, Z.~Dong, K.~Kong, I.M.~Lewis, Y.~Zhang and Y.-J.~Zheng,
  \emph{{Probing the CP structure of the top quark Yukawa at the future muon
  collider}}, \href{https://doi.org/10.1007/JHEP05(2024)176}{\emph{JHEP}
  {\bfseries 05} (2024) 176}
  [\href{https://arxiv.org/abs/2311.07645}{{\ttfamily 2311.07645}}].

\bibitem{Dedes:2017zog}
A.~Dedes, W.~Materkowska, M.~Paraskevas, J.~Rosiek and K.~Suxho, \emph{{Feynman
  rules for the Standard Model Effective Field Theory in R$_{\xi}$ -gauges}},
  \href{https://doi.org/10.1007/JHEP06(2017)143}{\emph{JHEP} {\bfseries 06}
  (2017) 143} [\href{https://arxiv.org/abs/1704.03888}{{\ttfamily
  1704.03888}}].

\bibitem{Chen:2021rid}
J.~Chen, C.-T.~Lu and Y.~Wu, \emph{{Measuring Higgs boson self-couplings with 2
  \textrightarrow{} 3 VBS processes}},
  \href{https://doi.org/10.1007/JHEP10(2021)099}{\emph{JHEP} {\bfseries 10}
  (2021) 099} [\href{https://arxiv.org/abs/2105.11500}{{\ttfamily
  2105.11500}}].

\bibitem{Chen:2021pqi}
J.~Chen, T.~Li, C.-T.~Lu, Y.~Wu and C.-Y.~Yao, \emph{{Measurement of Higgs
  boson self-couplings through 2\textrightarrow{}3 vector bosons scattering in
  future muon colliders}},
  \href{https://doi.org/10.1103/PhysRevD.105.053009}{\emph{Phys. Rev. D}
  {\bfseries 105} (2022) 053009}
  [\href{https://arxiv.org/abs/2112.12507}{{\ttfamily 2112.12507}}].

\bibitem{Alwall:2014hca}
J.~Alwall, R.~Frederix, S.~Frixione, V.~Hirschi, F.~Maltoni, O.~Mattelaer
  et~al., \emph{{The automated computation of tree-level and next-to-leading
  order differential cross sections, and their matching to parton shower
  simulations}}, \href{https://doi.org/10.1007/JHEP07(2014)079}{\emph{JHEP}
  {\bfseries 07} (2014) 079} [\href{https://arxiv.org/abs/1405.0301}{{\ttfamily
  1405.0301}}].

\bibitem{Degrande:2020evl}
C.~Degrande, G.~Durieux, F.~Maltoni, K.~Mimasu, E.~Vryonidou and C.~Zhang,
  \emph{{Automated one-loop computations in the SMEFT}},
  \href{https://arxiv.org/abs/2008.11743}{{\ttfamily 2008.11743}}.

\bibitem{Ballestrero:2020qgv}
A.~Ballestrero, E.~Maina and G.~Pelliccioli, \emph{{Different polarization
  definitions in same-sign $WW$ scattering at the LHC}},
  \href{https://doi.org/10.1016/j.physletb.2020.135856}{\emph{Phys. Lett. B}
  {\bfseries 811} (2020) 135856}
  [\href{https://arxiv.org/abs/2007.07133}{{\ttfamily 2007.07133}}].

\bibitem{De:2020iwq}
S.~De, V.~Rentala and W.~Shepherd, \emph{{Measuring the polarization of
  boosted, hadronic $W$ bosons with jet substructure observables}},
  \href{https://arxiv.org/abs/2008.04318}{{\ttfamily 2008.04318}}.

\end{thebibliography}\endgroup

\end{document}